\documentclass[runningheads]{llncs}
\usepackage{graphicx}
\usepackage{subfig} 
\usepackage[T1]{fontenc}
\usepackage{graphicx}
\usepackage{amsmath,amssymb,amsfonts}
\usepackage{booktabs}
\usepackage{multirow}
\usepackage{xcolor}
\usepackage{algorithm}
\usepackage{algorithmic}
\usepackage{float}
\begin{document}

\title{MelShield: Robust Mel-Domain Audio Watermarking for Provenance Attribution of AI Generated Synthesized Speech}
\titlerunning{MelShield}

\author{Yutong Jin\inst{1} \and Qi Li\inst{1} \and Lingshuang Liu\inst{2} \and Jianbing Ni\inst{1}}
\authorrunning{Y. Jin et al.}

\institute{
Department of Electrical and Computer Engineering, Queen's University, Kingston, ON, Canada \\
\email{23dc4@queensu.ca, qi.li@queensu.ca, jianbing.ni@queensu.ca}
\and
Department of Electrical and Computer Engineering, University of Waterloo, Waterloo, ON, Canada
\email{lingshaung.liu@uwaterloo.ca}
}
\maketitle

\begin{abstract}
In this paper, we propose MelShield, a robust, in-generation, keyed audio watermarking framework that embeds identifiable signals into AI-generated audio for copyright protection and reliable attribution. Specifically, MelShield operates in the Mel-spectrogram domain during the generation process, targeting intermediate acoustic representations in Mel-conditioned pipelines for text-to-speech (TTS) generation. 
The core idea is to treat the intermediate Mel-spectrogram as the host signal and embed a short binary payload via low-energy, keyed spread-spectrum perturbations distributed across carefully selected time–frequency regions prior to waveform synthesis. By performing watermarking before vocoder inference, MelShield remains plug-and-play for Mel-conditioned TTS architectures and does not require modification or retraining of the underlying TTS generation vocoder, such as DiffWave and HiFi-GAN. Moreover, the multi-user keyed construction enables scalable user-specific attribution, while the keyed verification mechanism limits unauthorized decoding, thereby reducing the risk of large-scale extractor probing and adversarial analysis.
Extensive experiments on DiffWave and HiFi-GAN demonstrate that MelShield achieves reliable watermark extraction, approaching 100\% bit accuracy, even under signal distortions, e.g., compression and additive noise, while preserving high perceptual audio quality.


\keywords{Audio watermarking \and Text to speech generation \and Audio processing \and Copyright protection and attribution \and Neural vocoders }
\end{abstract}

\section{Introduction}


Neural text-to-speech (TTS) systems have advanced rapidly in recent years, enabling large-scale generation of highly natural and intelligible synthesized speech~\cite{ren2020fastspeech,shen2018natural,van2016wavenet}. With only a short enrollment recording, generative models can create speech that is perceptually indistinguishable from natural human speech~\cite{jia2018transfer}. These technologies are widely deployed across a broad range of applications, including conversational agents, audiobook narration, accessibility services, virtual assistants, and media production pipelines.
Most TTS architectures adopt a two-stage generation pipeline, with representative systems including Glow-TTS~\cite{kim2020glowtts}, FastSpeech2 \cite{ren2020fastspeech}, and Tacotron2 \cite{shen2018natural}. In the first stage, a front-end acoustic model predicts an intermediate representation, typically a Mel-spectrogram, from textual input. In the second stage, a neural vocoder synthesizes the final waveform from this representation. Neural vocoders are implemented using diverse generative paradigms, such as diffusion-based approaches (e.g., DiffWave~\cite{kong2020diffwave}) and generative adversarial networks (GAN) (e.g., HiFi-GAN~\cite{kong2020hifi}). However, the increasing realism and accessibility of speech synthesis technologies also amplify concerns regarding misuse, provenance, and copyright protection. In particular, audio deepfakes can convincingly imitate target speakers and deliver arbitrary content, creating serious risks related to misinformation, impersonation, fraud, and unauthorized redistribution~\cite{klein2024source}. These challenges underscore the urgent need for practical mechanisms that enable reliable tracing and attribution of AI-generated speech to its source.

Audio watermarking has emerged as a proactive mechanism for provenance attribution in AI-generated audio~\cite{liu2025xattnmark}. Rather than relying solely on post hoc detection of synthetic content, watermarking embeds a traceable signal during or around the generation process, enabling ownership or origin to be verified through a dedicated extractor. This proactive embedding paradigm strengthens accountability by ensuring that attribution information is intrinsically associated with the generated audio.
In the context of generative audio systems, existing watermarking approaches are generally categorized into two paradigms: post-hoc watermarking and in-generation watermarking.

\textbf{Related Work}.
Post-hoc watermarking methods embed watermarks after waveform synthesis. Recent approaches, including WavMark~\cite{chen2023wavmark}, AudioSeal~\cite{roman2024proactive}, and Timbre Watermarking~\cite{liu2023detecting}, demonstrate strong robustness and high perceptual fidelity under a variety of signal distortions. Nevertheless, from the standpoint of practical copyright attribution and large-scale deployment, several limitations remain~\cite{wen2025sok,li2024proactive}.
First, because watermark embedding is performed as a separate processing stage after generation, the watermarking module must operate alongside the generative system. This architectural separation introduces an explicit post-processing component that may be bypassed, disabled, or weakened during real-world deployment or downstream transformations. Second, payload capacity is typically limited, often on the order of tens of bits (e.g., 32-bit messages), which constrains scalability when distinct identifiers are required for a large user base in multi-user copyright distribution settings. Third, many post-hoc schemes rely on public or broadly accessible extractors. While convenient for open verification, this design also facilitates large-scale probing by adversaries, enabling systematic extractor analysis and the development of targeted removal or evasion attacks.

{In-generation} watermarking has emerged as a promising paradigm in which the watermark is embedded directly within the generative process. By integrating watermark injection into model inference, this approach strengthens robustness and increases the practical difficulty of removal. A representative example is GROOT~\cite{liu2024groot}, which embeds watermark information into the initial noise of a diffusion-based audio generation model using a convolutional neural network and employs a neural extractor for recovery. Although GROOT achieves improved robustness and higher payload capacity, it presents practical constraints: it is limited to diffusion-based architectures, introduces non-negligible computational overhead, and may degrade audio fidelity under certain configurations. Moreover, if the extractor is publicly accessible, the system remains vulnerable to large-scale probing and reverse engineering. Another representative system, TraceableSpeech~\cite{zhou2024traceablespeech}, integrates watermarking into a VALL-E-style~\cite{wang2023valle} discrete codec-based TTS pipeline and reports strong perceptual quality and imperceptibility. However, its design is tightly coupled to codec-token generation and does not readily generalize to diffusion-based or GAN-based vocoders.
Therefore, it is essential to design a lightweight, model-agnostic, and keyed in-generation watermarking framework that supports scalable multi-user attribution while remaining compatible with diverse generative paradigms.

\textbf{Contributions.} In this paper, we propose MelShield, a robust in-generation audio watermarking framework that operates directly in the Mel-spectrogram domain of vocoder-based speech synthesis pipelines. Our key insight is that the Mel-spectrogram functions as a shared, model-agnostic interface across a broad range of neural TTS architectures. Thus, MelShield can be seamlessly integrated into diverse generative models whenever a Mel representation appears as an intermediate acoustic feature, without requiring modifications to the underlying vocoder. Furthermore, the designed low-energy structured perturbations in the Mel domain can reliably survive waveform synthesis while preserving perceptual audio quality.
MelShield adopts a keyed watermarking design, ensuring that only a legitimate owner possessing the secret key with stored reference mel spectrogram can recover the embedded message. Verification can be performed on the user or service side using the stored reference Mel representation generated during synthesis, eliminating the need to train or fine-tune additional decoding models. This lightweight verification strategy reduces the risk of large-scale probing, extractor misuse, and adversarial reverse engineering, while remaining practical for copyright attribution scenarios.

Extensive experiments on both diffusion-based and GAN-based vocoders demonstrate that MelShield achieves strong robustness under common signal processing distortions while maintaining high perceptual quality. Notably, under practical compression settings such as MP3-128 and AAC-96, MelShield achieves perfect decoding accuracy (1.00), and under additive noise at 20dB, it maintains decoding accuracy above 95\%. The results further illustrate favorable trade-offs among audio fidelity, decoding reliability, and payload capacity, validating the practicality and generality of the proposed Mel-domain watermarking design for AI-generated speech attribution.

The main advantages of MelShield are summarized as follows.

\begin{itemize}
\item \textbf{Keyed in-generation Mel-domain watermarking.} MelShield embeds user-specific binary payloads directly into log-Mel spectrograms via low-energy, keyed spread-spectrum perturbations prior to waveform synthesis, ensuring seamless integration into the acoustic generation process.

\item \textbf{Retraining-free and model-agnostic deployment.} MelShield requires neither retraining nor architectural modification of the underlying audio generation vocoder and seamlessly integrates with a wide range of Mel-conditioned speech synthesis pipelines, including both diffusion-based and GAN-based vocoders, enabling practical and flexible deployment.

\item \textbf{Multi-user attribution.} By assigning distinct message–key pairs to individual users and enabling verification through user-side reference Mel representations, MelShield supports scalable multi-user copyright attribution while mitigating risks of unauthorized decoding and watermark forgery.

\item \textbf{Robustness-fidelity-capacity trade-off.} MelShield preserves high perceptual audio quality while achieving reliable decoding under common signal processing distortions and offers favorable trade-offs among audio fidelity, decoding robustness, and payload capacity, validating its practicality for real-world AI-generated speech attribution.
\end{itemize}

\section{Preliminaries}
In this section, we review the preliminaries of MelShield, including TTS pipelines, neural vocoders, audio watermarking, and Mel-spectrogram. 

\subsection{TTS Pipelines and Neural Vocoders}

Modern TTS systems synthesize speech from textual input using diverse neural architectures. Although certain models are designed in an end-to-end fashion, a widely adopted paradigm follows a two-stage pipeline: a text-to-acoustic model first maps linguistic inputs, such as text tokens, phonemes, or aligned linguistic features, into an intermediate acoustic representation, typically a Mel-spectrogram; subsequently, a neural vocoder converts this representation into a time-domain waveform~\cite{ren2020fastspeech,shen2018natural}.
Early neural vocoders were predominantly autoregressive (AR), such as WaveNet~\cite{van2016wavenet}, which achieved high-fidelity synthesis but incurred substantial inference latency due to sample-by-sample waveform generation. Recently, non-autoregressive architectures, particularly GAN-based and diffusion-based vocoders (e.g., HiFi-GAN~\cite{kong2020hifi}, DiffWave~\cite{kong2020diffwave}, and BigVGAN~\cite{lee2022bigvgan}), have become dominant in modern TTS systems. These models significantly improve generation efficiency while maintaining strong perceptual quality, offering different trade-offs between robustness, speed, and computational cost across architectural families.
In these variations, the intermediate Mel-spectrogram remains a widely adopted interface in contemporary pipelines. It provides a compact and perceptually meaningful representation of speech while maintaining compatibility across heterogeneous vocoder backends, thereby serving as a stable and model-agnostic acoustic abstraction.

Let $x$ denote the textual or linguistic input and $X \in \mathbb{R}^{C \times M}$ denote the predicted Mel-spectrogram with $C$ Mel bands and $M$ time frames. The front-end acoustic model can be abstractly written as $X = g_{\phi}(x),$ where $g_{\phi}$ is a neural text-to-spectrogram model with parameters $\phi$. A neural vocoder then synthesizes the waveform
$\hat{y} = f_{\theta}(X)$, where $\hat{y}$ is the generated speech waveform and $f_{\theta}$ denotes the spectrogram-to-waveform mapping. In this paper, we focus on watermarking the shared intermediate representation $X$ before waveform synthesis, while treating the vocoder as a black-box module.

\paragraph{Diffusion-based vocoders.}
Diffusion-based vocoders, such as DiffWave~\cite{kong2020diffwave}, generate audio by iteratively denoising a noisy waveform while conditioning on a Mel-spectrogram. Let $\{ \beta_t \}_{t=1}^{T_d}$ be a variance schedule and $y_0$ be a clean waveform sample. The forward diffusion process gradually perturbs the waveform by adding Gaussian noise:
\begin{equation}
q(y_t \mid y_{t-1}) =
\mathcal{N}\!\left(\sqrt{1-\beta_t}\,y_{t-1},\, \beta_t \mathbf{I}\right),
\quad t=1,\ldots,T_d.
\label{eq:diff-forward}
\end{equation}
At inference time, generation starts from Gaussian noise and applies a learned reverse process conditioned on the Mel-spectrogram:
\begin{equation}
y_{t-1} = h_{\theta}(y_t, t, X),
\label{eq:diff-reverse}
\end{equation}
where $h_{\theta}$ is a denoising network that predicts and removes noise at each step. After $T_d$ denoising steps, the final sample is taken as the synthesized waveform $\hat{y}$.

\paragraph{GAN-based vocoders.}
GAN-based vocoders, such as HiFi-GAN~\cite{kong2020hifi}, synthesize waveforms using a generator trained adversarially against one or more discriminators. Given a Mel-spectrogram $X$, the generator produces a waveform estimate
\begin{equation}
\hat{y} = G_{\theta}(X).
\label{eq:gan-vocoder}
\end{equation}
The generator is trained jointly with discriminator(s) $D_{\psi}$ under an adversarial objective, commonly combined with auxiliary reconstruction losses (e.g., feature matching or multi-resolution STFT losses) to improve perceptual fidelity. A simplified conditional adversarial objective can be written as
\begin{equation}
\min_{\theta}\max_{\psi}
\ \mathbb{E}_{(y,X)}[\log D_{\psi}(y,X)]
+\mathbb{E}_{X}[\log(1-D_{\psi}(G_{\theta}(X),X))].
\label{eq:gan-objective}
\end{equation}
Compared to diffusion vocoders, GAN-based vocoders are typically faster at inference as synthesis is obtained in a single forward pass, while diffusion vocoders often provide a different robustness-quality trade-off due to iterative denoising.

\paragraph{Relevance to this work.}
Both diffusion-based and GAN-based vocoders commonly adopt Mel-spectrograms as conditioning inputs, making the Mel domain a natural and model-agnostic insertion point for watermark embedding. This shared acoustic interface allows a unified watermarking mechanism to be seamlessly integrated across diverse TTS pipelines, without requiring retraining or architectural modifications to the underlying vocoder.

\subsection{Audio Watermarking}
\label{subsec:audio-watermarking}

Audio watermarking aims to embed a hidden message into an audio signal while preserving perceptual quality and enabling reliable recovery at a later stage. Let $x$ denote a clean audio signal and let $\mathbf{m} \in \{0,1\}^{L}$ denote an $L$-bit binary message to be embedded. A watermarking system typically consists of two components: an embedding function $\mathcal{E}$ and an extraction function $\mathcal{D}$. The embedder generates a watermarked audio signal
\vspace{-0.05in}
\begin{equation}
x_w = \mathcal{E}(x,\mathbf{m}),
\label{eq:wm_embed_general}
\end{equation}
The extractor attempts to recover the embedded message from the watermarked audio (or a transformed version of it) as
\vspace{-0.05in}
\begin{equation}
\hat{\mathbf{m}} = \mathcal{D}(x_w),
\label{eq:wm_extract_general}
\end{equation}
or more generally from a received signal $\tilde{x}$,
\vspace{-0.05in}
\begin{equation}
\hat{\mathbf{m}} = \mathcal{D}(\tilde{x}).
\label{eq:wm_extract_received}
\end{equation}
Ideally, the recovered message matches the embedded one, i.e., $\hat{\mathbf{m}}=\mathbf{m}$. In practice, watermark recovery is evaluated in terms of a decoding accuracy threshold, since channel distortions and post-processing may introduce bit errors. We therefore require the recovered payload to satisfy a minimum bit-wise accuracy, and treat successful decoding as $\mathrm{BitAcc}(\mathbf{m},\hat{\mathbf{m}})\ge\tau_{\mathrm{acc}}$, where $\tau_{\mathrm{acc}}$ is an application-dependent threshold and the ideal case corresponds to $\tau_{\mathrm{acc}}=1.0$.

We quantify watermark recovery performance using bit-wise decoding accuracy. For an $L$-bit payload $\mathbf{m}\in\{0,1\}^{L}$ and a decoded message $\hat{\mathbf{m}}=\mathcal{D}(\tilde{x},K)$, BitAcc is defined as
\vspace{-0.05in}
\begin{equation}
\mathrm{BitAcc}(\mathbf{m},\hat{\mathbf{m}})
=\frac{1}{L}\sum_{i=1}^{L}\mathbb{I}\!\left[\hat{m}_i=m_i\right],
\label{eq:bitacc_def}
\vspace{-0.1in}
\end{equation}
where $\mathbb{I}[\cdot]$ is the indicator function. In our experiments, we report BitAcc under different channel distortions and operating points.

A secure audio watermarking system should satisfy several core properties.
\vspace{-0.1in}
\paragraph{Imperceptibility (Fidelity).}
Watermark embedding should not introduce objectionable audible distortion. Let
$q(\cdot,\cdot)$ denote a perceptual quality metric in which larger values indicate
better quality (e.g., PESQ). Imperceptibility can be expressed as
\begin{equation}
q(x,x_w) \ge \tau_q,
\label{eq:wm_imperceptibility}
\end{equation}
where $\tau_q$ is an application-dependent minimum quality threshold. In
experimental evaluation, this property is typically assessed using perceptual and
intelligibility metrics such as DNSMOS~\cite{reddy2021dnsmos}, PESQ~\cite{rix2001perceptual}, and STOI~\cite{taal2011stoi}.

\vspace{-0.1in}
\paragraph{Robustness.}
The embedded watermark should remain decodable after common signal processing operations such as compression, resampling, filtering, reverberation, and additive noise. Let $\mathcal{T}(\cdot)$ denote such a transformation applied to the watermarked audio. In our evaluation, robustness is defined jointly over decoding reliability and perceptual quality. For a target quality metric $q(\cdot,\cdot)$ (e.g., PESQ, STOI, or DNSMOS) and thresholds $\tau_q$ and $\tau_{\mathrm{acc}}$, we require
\begin{equation}
q\!\left(x,\mathcal{T}(x_w)\right)\ge\tau_q
\quad \wedge \quad
\mathrm{BitAcc}\!\left(\mathbf{m},\mathcal{D}(\mathcal{T}(x_w))\right)\ge\tau_{\mathrm{acc}}.
\label{eq:wm_robustness_q_acc}
\end{equation}
In practice, we sweep $\mathcal{T}$ over a set of attacks and measure both the resulting quality metrics and bit accuracy.

\paragraph{Key-based security.}
MelShield is a keyed watermarking scheme: successful recovery should require the correct key, while decoding with an incorrect key should behave similarly to the no-watermark case. Let $S(\tilde{x},K)$ denote a verification statistic computed under key $K$ (e.g., bit-wise accuracy against an expected message, or an aggregate correlation score), and let $\tau$ be a decision threshold. 
\begin{equation}
S(\tilde{x},K)\ge\tau \ \text{for}\ \tilde{x}=\mathcal{T}(x_w),
\label{eq:sec_correct_key}
\end{equation}
while for any mismatched key $K'\neq K$,
\vspace{-0.05in}
\begin{equation}
S(\tilde{x},K')<\tau,
\label{eq:sec_wrong_key}
\end{equation}
with high probability, and the distribution of $S(\tilde{x},K')$ is close to that obtained under $H_0$ (no watermark). This captures that unauthorized parties without the correct key cannot reliably decode or pass verification for a target identity.

\vspace{-0.1in}
\paragraph{Capacity.}
Capacity measures how much information can be embedded in the host signal, typically quantified by the payload length $L$ (in bits) or by a bitrate normalized by audio duration. In practice, capacity is coupled with fidelity and robustness: increasing payload usually increases embedding strength, which may degrade perceptual quality or reduce decoding reliability under attacks.

Overall, practical audio watermarking is governed by a multi-objective trade-off among \emph{fidelity}, \emph{robustness}, \emph{security}, and \emph{capacity}. The goal of a well-designed system is not to optimize only one of these dimensions, but to achieve a favorable operating point for the target application. In this work, we focus on an in-generation watermarking design that targets strong robustness and multi-user attribution while maintaining high audio fidelity and scalable payload capacity.

\subsection{Mel-Spectrogram}
\label{subsec:mel_spectrogram}

The Mel scale is a perceptual frequency scale designed to better reflect human pitch perception, originally introduced in psychophysical studies of pitch~\cite{stevens1937mel}. Given a discrete-time waveform $y[n]$, a Mel-spectrogram is computed by first applying a short-time Fourier transform (STFT) and then projecting the resulting magnitude or power spectrum onto a bank of Mel-spaced triangular filters. Concretely, for frame index $t$ and frequency bin $k$, the STFT is
\vspace{-0.1in}
\begin{equation}
Y(t,k)=\sum_{n=0}^{N-1} y[n+tH]\;w[n]\;e^{-j2\pi kn/N},
\label{eq:stft}
\end{equation}
where $w[n]$ is a window function, $N$ is the FFT size, and $H$ is the hop size. Let $P(t,k)=|Y(t,k)|^{p}$ denote the magnitude or power spectrum with exponent $p\in\{1,2\}$. A Mel filterbank $\{b_c(k)\}_{c=1}^{C}$ then aggregates linear-frequency bins into $C$ Mel bands,
\vspace{-0.1in}
\begin{equation}
X(t,c)=\sum_{k} b_c(k)\,P(t,k), \quad c=1,\dots,C,
\label{eq:mel_filterbank}
\end{equation}
followed by a log compression or normalization step to obtain the log-Mel representation
\begin{equation}
\tilde{X}(t,c)=\log\!\bigl(X(t,c)+\epsilon\bigr),
\label{eq:log_mel}
\end{equation}
where $\epsilon>0$ is a small constant for numerical stability.

In many TTS pipelines, the Mel-spectrogram is the intermediate representation between the text-to-acoustic model and the neural vocoder. This shared interface motivates embedding watermarks directly in the Mel domain before waveform synthesis, enabling integration across diverse Mel-conditioned TTS systems.

\section{Problem Formulation}
\label{sec:threat_model}

We consider a provenance attribution scenario for neural TTS systems in which the AI-generated audio may be distributed through untrusted channels, such as social media platforms, messaging applications, or content-sharing services. To achieve copyright protection, during synthesis, the legitimate owner, such as an end user, platform operator, or model provider, embeds a user-specific or model-specific watermark into the intermediate Mel-spectrogram. This watermarked Mel representation is subsequently used by the neural vocoder to generate the final speech waveform, ensuring that each produced audio clip carries a unique and traceable identifier.

Our goal is to enable the owner to verify whether a suspect audio sample originates from a particular synthesis instance, even after common downstream transformations. These transformations may include compression, resampling, amplitude scaling, additive noise, reverberation, or other typical signal processing distortions encountered in real-world distribution. Given a suspect waveform, the verifier should reliably recover the expected watermark message when the audio is derived from the protected output. In such cases, the extracted watermark should achieve high bit accuracy with respect to the originally embedded message, indicating correct attribution. Conversely, the use of mismatched keys or unrelated audio should result in decoding failure or incorrect message recovery with high probability.

We consider a reference-based verification setting in which the legitimate owner retains the reference Mel-spectrogram generated during synthesis for subsequent attribution. This assumption is practical in both user-side and service-side TTS deployments, as the generating party inherently has access to the intermediate Mel representation at the time of creation.

\begin{figure}[t]
    \centering
    \includegraphics[width=0.8\linewidth]{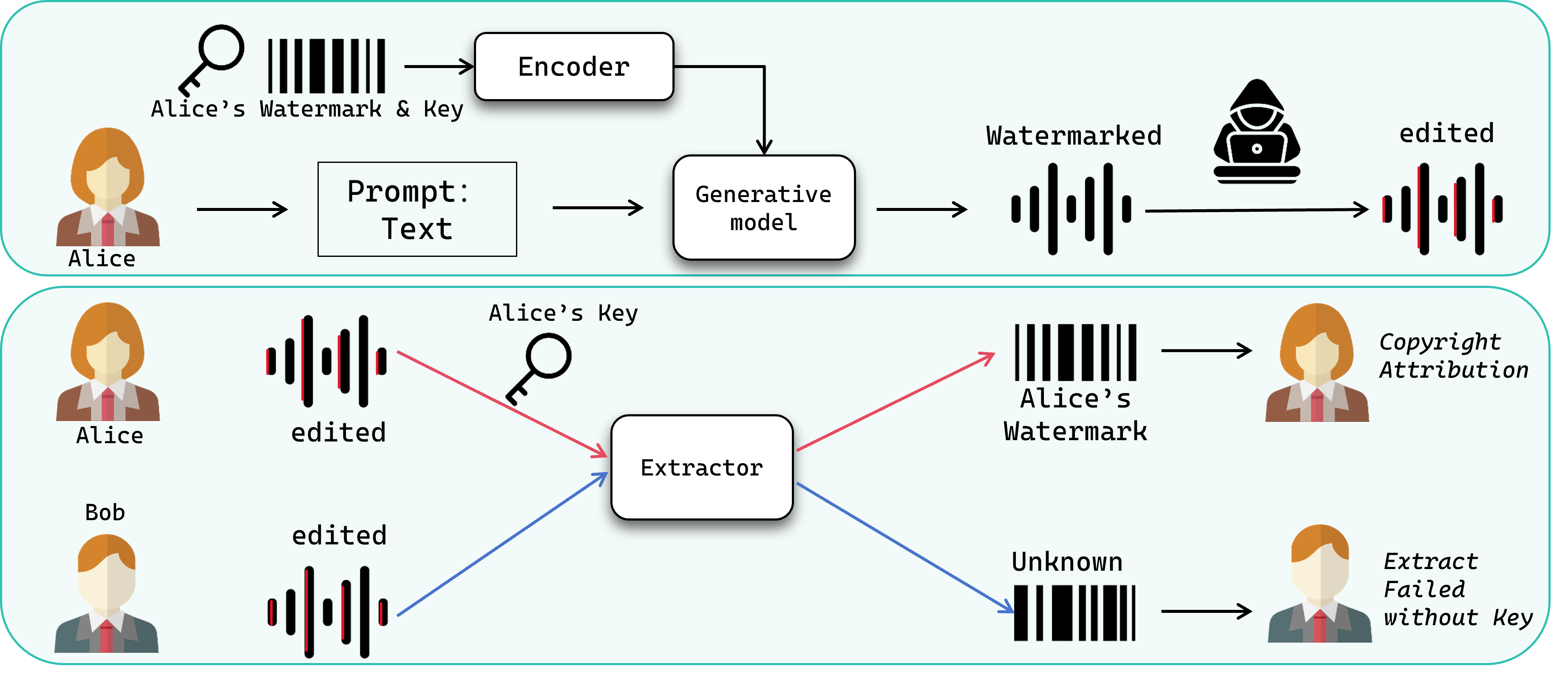}
    \caption{Workflow of keyed audio watermark embedding and owner-side verification for copyright attribution.}
    \label{fig:overview}
\end{figure}

\vspace{-0.1in}
\paragraph{Security and Performance Objectives.}
MelShield is designed for threshold-based attribution using bit-wise decoding accuracy. Let $\hat{\mathbf{m}}$ be the recovered payload under the claimed identity and key, and let $\mathrm{BitAcc}(\mathbf{m},\hat{\mathbf{m}})$ be defined as in Eq.~\eqref{eq:bitacc_def}. For a verification threshold $\tau_{\mathrm{acc}}$, a sample is accepted as belonging to the claimed identity if $\mathrm{BitAcc}(\mathbf{m},\hat{\mathbf{m}})\ge\tau_{\mathrm{acc}}$. For correctly watermarked audio generated under the claimed key, the verifier should recover the payload with sufficiently high bit-wise accuracy and thus pass verification. Conversely, for unwatermarked audio, audio verified under an incorrect key, or audio in which the watermark has been effectively destroyed, the recovered payload should exhibit low bit-wise accuracy relative to the claimed message and fail verification with high probability.

\begin{itemize}

\item \textbf{Verifiability:} For protected outputs, the embedded watermark should be verifiable when the recovered payload satisfies $\mathrm{BitAcc}(\mathbf{m},\hat{\mathbf{m}})\ge\tau_{\mathrm{acc}}$ under typical distribution-channel distortions.

  \item \textbf{Robustness:} The embedded watermark should remain verifiable after common downstream processing operations, including lossy compression, resampling, moderate additive noise, amplitude scaling, reverberation, and simple edits that largely preserve speech content.

  \item \textbf{Perceptual transparency:} Embedding should introduce only low-energy, minimally perceptible perturbations, ensuring that synthesized speech quality remains comparable to unwatermarked generation.

  \item \textbf{Keyed multi-user attribution:} Distinct message--key pairs are assigned to different users or deployment instances; without the correct key, an adversary cannot reliably decode, impersonate, or forge a valid watermark for a target identity, and verification should fail with high probability.
\end{itemize}

Fig.~\ref{fig:overview} illustrates a representative workflow. During synthesis, a user-side or service-side generator embeds a keyed watermark into the intermediate Mel representation and then produces audio via a neural vocoder (e.g., HiFi-GAN or DiffWave). Later, the legitimate owner uses the corresponding secret key to verify a suspect audio clip for copyright attribution, while unauthorized parties without the key cannot reliably validate the claimed watermark.

\section{MelShield}
\label{sec:method}

\begin{figure}[t]
    \centering
    \includegraphics[width=0.8\linewidth]{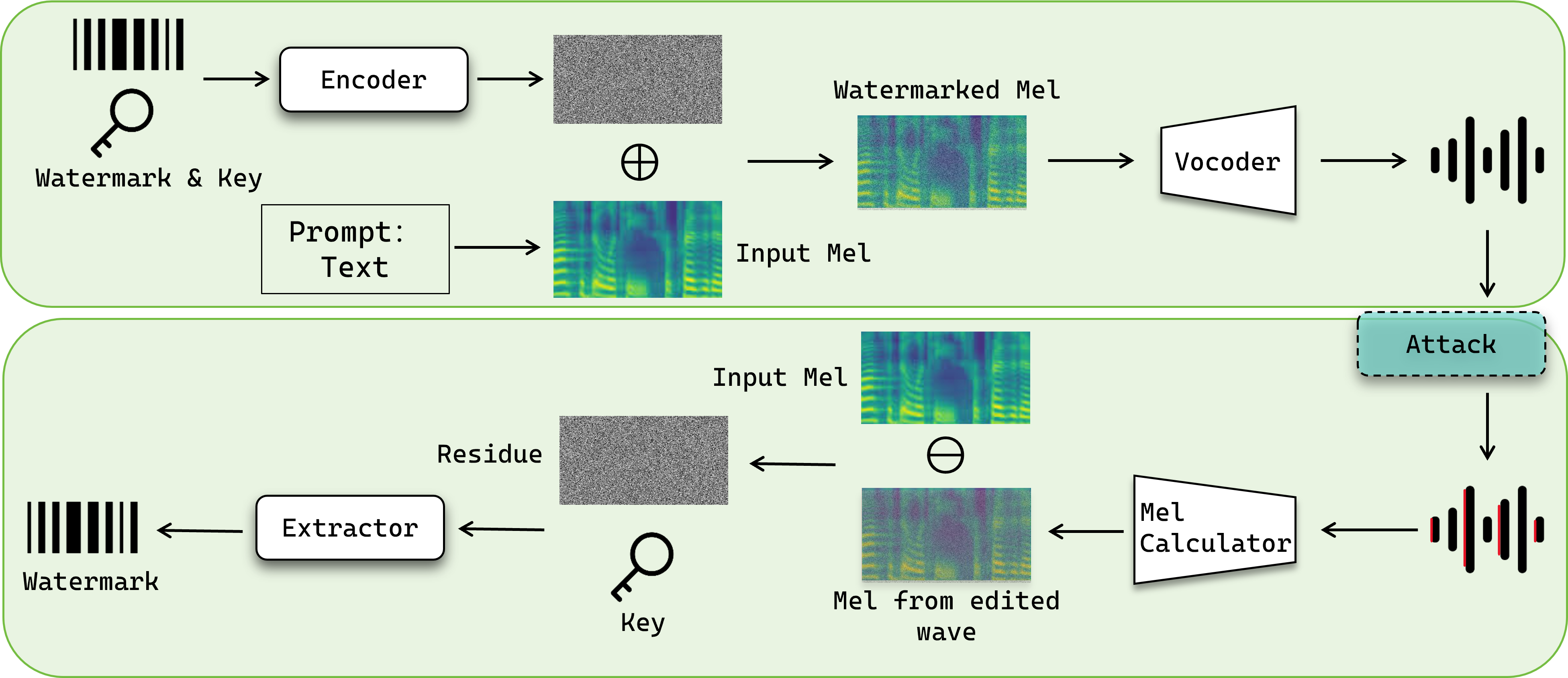}
    \caption{Workflow of Audio Watermark Embedding and Verification}
    \label{fig:method-overview}
    \vspace{-0.1in}
\end{figure}

In this section, we present MelShield, the keyed in-generation watermarking method in the log-Mel spectrogram domain.
Fig.~\ref{fig:method-overview} gives an overview: a short
binary message is expanded into pseudorandom $\{-1,1\}$ patterns, added as a low-energy perturbation to the selected time-frequency bins of the Mel-spectrogram, and then passed through an off-the-shelf vocoder.  
During verification, the verifier reconstructs a Mel-spectrogram from a suspect waveform, compares it with a stored reference Mel, and recovers the message by a keyed correlation test.

To formalize this process, let $X \in [0,1]^{C \times M}$ denote the unwatermarked normalized log-Mel spectrogram produced by a TTS front-end, with $C$ Mel bands and $M$ time frames. A vocoder then synthesizes the waveform $\hat{y} = f_\theta(X)$, where $f_\theta$ is either a diffusion- or GAN-based vocoder. Our method introduces small, structured perturbations to the original Mel-spectrogram $X$, producing a watermarked representation $\tilde{X}$. The neural vocoder then operates on the modified input, generating the waveform $\hat{y} = f_\theta(\tilde{X})$.

\subsection{Watermark Embedding}
\label{subsec:generation}
Given an unwatermarked normalized log-Mel spectrogram $X$, the embedding stage constructs a keyed perturbation over a selected mid-frequency region and adds it to $X$ with adaptive energy control, yielding the watermarked spectrogram $\tilde{X}$. A user (or deployment instance) is assigned a binary payload $\mathbf{m}=(m_1,\dots,m_L)\in\{0,1\}^L$ and a secret key $K$. The payload specifies the attribution information (e.g., a user ID or service tag), while $K$ seeds all pseudorandom choices used during embedding.

We embed only in a mid-frequency band $\mathcal{F}=\{c_{\min},\dots,c_{\max}-1\}$ across all time frames, which empirically provides a favorable robustness--transparency trade-off. Let $X_{\mathcal{F}}\in[0,1]^{|\mathcal{F}|\times M}$ denote the restriction of $X$ to this band. To encode an $L$-bit payload, we generate $L$ key-dependent spreading patterns $S_j^{(K)}=\mathsf{Gen}(K,j;|\mathcal{F}|,M)\in\{-1,1\}^{|\mathcal{F}|\times M}$. Each $S_j^{(K)}$ is a full-region $\pm1$ pattern reproducible from $(K,j)$. We map each bit to a polarity $d_j=2m_j-1\in\{-1,+1\}$, so that the $j$-th bit selects the sign of its spreading layer $d_jS_j^{(K)}$. We then superpose all bit layers into a single watermark layer
\begin{equation}
W=\frac{1}{\sqrt{L}}\sum_{j=1}^{L} d_j S_j^{(K)},
\end{equation}
where the factor $1/\sqrt{L}$ keeps the perturbation energy approximately stable as the payload length varies.

To improve fidelity, we compute an adaptive mask $A\in[0,1]^{|\mathcal{F}|\times M}$ from the clean region $X_{\mathcal{F}}$ by estimating a frame-level energy weight and broadcasting it along the frequency axis. This mask downweights embedding in low-energy frames and mitigates clipping near the normalization boundaries, thereby reducing audible artifacts. With embedding strength $\alpha>0$, watermark injection is performed only within $\mathcal{F}$ via
\begin{equation}
\tilde{X}_{\mathcal{F}}=\operatorname{clip}_{[0,1]}\!\bigl(X_{\mathcal{F}}+\alpha (A\odot W)\bigr),
\end{equation}
where $\odot$ denotes element-wise multiplication (Hadamard product), while all time--frequency bins outside $\mathcal{F}$ remain unchanged. In practice, we apply a small headroom margin before embedding to further reduce destructive clipping near 0 and 1.

For owner-side verification, the encoder stores a compact reference copy $X_{\text{ref}}$ of the original Mel spectrogram together with meta-data (e.g., $L$, $\alpha$, Mel dimensions, $\mathcal{F}$, and a key identifier). The watermarked spectrogram $\tilde{X}$ is then passed to the vocoder for waveform synthesis as in $\hat{y} = f_\theta(\tilde{X})$.

\subsection{Watermark Extraction}
\label{subsec:extraction}
Given a suspect waveform $\tilde{y}$, the legitimate owner can verify whether it originates from a protected utterance and, if so, recover the embedded payload. The verifier holds the secret key $K$ and a reference record for the candidate utterance, including the stored log-Mel spectrogram $X_{\text{ref}}$ and the associated meta-data (e.g., $\mathcal{F}$, $L$, and $\alpha$). The verifier reconstructs the suspect log-Mel spectrogram using the same front-end transform $X_{\text{det}}=\mathcal{T}(\tilde{y})$ and aligns it with $X_{\text{ref}}$ in time to obtain aligned arrays of shape $C\times M'$, where $M'$ denotes the number of aligned time frames and equal to the original frame length $M$. The verifier then computes a reference-based residual
\begin{equation}
\Delta = X_{\text{det}} - X_{\text{ref}},\qquad
\Delta \leftarrow \Delta - \frac{1}{CM'}\sum_{c,t}\Delta(c,t),
\label{eq:residual}
\end{equation}
where the mean subtraction removes global offsets induced by gain changes or normalization drift.

Using the stored meta-data and key $K$, the verifier regenerates the embedding band $\mathcal{F}$, recomputes the same adaptive mask $A$ from the clean reference region $X_{\text{ref},\mathcal{F}}$, and regenerates the bit-specific spreading patterns $S_j^{(K)}=\mathsf{Gen}(K,j;|\mathcal{F}|,M')\in\{\pm1\}^{|\mathcal{F}|\times M'}$ for $j=1,\dots,L$. For each bit, we compute a masked correlation score over the embedding region:
\begin{equation}
s_j(K)=\bigl\langle \Delta_{\mathcal{F}},A\odot S_j^{(K)}\bigr\rangle=\sum_{(c,t)\in\mathcal{F}\times\{0,\dots,M'-1\}}\Delta(c,t)\,A(c,t)\,S_j^{(K)}(c,t).
\label{eq:score-corr}
\end{equation}
Under the correct key and a watermarked sample, the residual approximately follows $\Delta_{\mathcal{F}}\approx\alpha(A\odot W)+\text{noise}$, where $W=\frac{1}{\sqrt{L}}\sum_{j=1}^{L} d_j S_j^{(K)}$ and $d_j=2m_j-1$. Therefore, $s_j(K)$ tends to be positive for $m_j=1$ and negative for $m_j=0$. For unrelated audio or a wrong key, the regenerated $S_j^{(K)}$ is effectively uncorrelated with the residual, so $s_j(K)$ concentrates near zero.

The verifier decodes by the sign of the score,
\[
\hat{m}_j=\mathbb{I}[s_j(K)\ge 0],\quad j=1,\dots,L,
\]
and obtains the recovered payload $\hat{\mathbf{m}}$. The magnitudes $|s_j(K)|$ serve as confidence values and can be aggregated into a global watermark presence test.

\section{Experiments}
\label{sec:experiments}

\subsection{Experimental Setup}

\paragraph{Datasets and TTS pipelines.}
We conduct experiments on the LJSpeech 1.1 dataset~\cite{ito2017ljspeech}, which consists of 13100 1-10 seconds single-speaker English read speech sampled at 22.05~kHz. For waveform synthesis, we evaluate two representative Mel-conditioned vocoders: the official DiffWave implementation and a HiFi-GAN generator configured for 22.05~kHz single-speaker speech. For watermark embedding and verification, we compute a normalized log-Mel spectrogram using an STFT-based Mel filterbank (Mel-Spectrogram). Specifically, we set $f_{\min}=20$~Hz, $f_{\max}=\tfrac{1}{2}\mathrm{sr}$, and use $C=80$ Mel frequency bins. A logarithmic transformation is then applied to obtain the log-Mel representation, which serves as the input to the watermark encoder and decoder.

\begin{table}[t]
\centering
\caption{Fidelity and decoding performance under different payload capacities on DiffWave and HiFi-GAN. ``Benchmark'' denotes the unwatermarked generated audio.}
\label{tab:fidelity_capacity_combined}
\setlength{\tabcolsep}{3.2pt}
\scriptsize
\resizebox{\columnwidth}{!}{%
\begin{tabular}{llccccccccc}
\toprule
Vocoder & Metric & Benchmark & 16 & 32 & 48 & 64 & 96 & 128 & 256 & 512 \\
\midrule
\multirow{5}{*}{HiFi-GAN}
& PESQ ref GT  & 3.8488 & 3.8212 & 3.8132 & 3.8007 & 3.7900 & 3.7732 & 3.7701 & 3.6504 & 3.5959 \\
& PESQ ref BM & --     & 4.2946 & 4.1209 & 4.0893 & 4.0333 & 3.9801 & 3.9328 & 3.7480 & 3.5406 \\
& STOI         & 0.9960 & 0.9960 & 0.9915 & 0.9913 & 0.9890 & 0.9888 & 0.9832 & 0.9751 & 0.9600 \\
& MOS          & 3.7992 & 3.7635 & 3.7506 & 3.7487 & 3.7490 & 3.7350 & 3.7289 & 3.6711 & 3.6508 \\
& ACC          & --     & 1.0000 & 1.0000 & 1.0000 & 1.0000 & 1.0000 & 1.0000 & 1.0000 & 0.9965 \\
\midrule
\multirow{5}{*}{DiffWave}
& PESQ ref GT  & 3.7002 & 3.6773 & 3.6348 & 3.6009 & 3.5899 & 3.5387 & 3.5011 & 3.3501 & 3.2285 \\
& PESQ ref BM & --     & 3.6112 & 3.5762 & 3.5556 & 3.5338 & 3.5223 & 3.5020 & 3.3520 & 3.0883 \\
& STOI         & 0.9769 & 0.9744 & 0.9722 & 0.9690 & 0.9689 & 0.9687 & 0.9623 & 0.9584 & 0.9456 \\
& MOS          & 3.6570 & 3.6637 & 3.6518 & 3.6382 & 3.6373 & 3.6376 & 3.6244 & 3.6082 & 3.5611 \\
& ACC          & --     & 1.0000 & 1.0000 & 0.9983 & 0.9919 & 0.9854 & 0.9773 & 0.9572 & 0.9307 \\
\bottomrule
\end{tabular}%
}
\vspace{-0.05in}
\end{table}

\vspace{-0.1in}
\paragraph{Watermark configuration.}
We evaluate MelShield along three dimensions: perceptual fidelity, payload capacity, and robustness against signal distortions. For the fidelity-capacity analysis, we adopted model-specific embedding-strength ranges based on the intrinsic sensitivity of each vocoder. Specifically, $\alpha$ was varied from 0 to 0.05 for DiffWave and from 0 to 0.4 for HiFi-GAN, while payload lengths ranged from 16 to 1024 bits to characterize the trade-off between audio quality and embedding capacity. For robustness evaluation, we fix the payload length to $L=32$ bits and assign each utterance a unique 32-bit message. To ensure fair comparison, the same 32-bit message is used across all evaluated methods under a given model configuration. Unless otherwise specified, we use a fixed random seed $k$ and restrict watermark embedding to the mid-frequency Mel band $\mathcal{F}=\{20,\dots,55\}$ on log-Mel spectrograms normalized to $[0,1]$. We set $\alpha=0.025$ for DiffWave and $\alpha=0.25$ for HiFi-GAN to account for their differing sensitivities to Mel-domain perturbations. All spreading patterns are deterministically generated from the secret key $K$ and the utterance identifier, and are stored as part of the meta-data.

\subsection{Evaluation Metrics}
We evaluate watermark reliability using bit-wise decoding accuracy. Perceptual audio quality is assessed using PESQ, STOI, and MOS(DNSMOS). PESQ is an intrusive perceptual quality metric that compares a degraded signal against a clean reference waveform and is widely used for end-to-end speech quality assessment in telephony and speech processing applications~\cite{rix2001perceptual}. STOI is an intrusive intelligibility metric designed to estimate speech intelligibility under additive noise and time-frequency processing~\cite{taal2011stoi}. DNSMOS is a non-intrusive neural metric that predicts mean opinion score (MOS)-like perceptual quality without requiring a clean reference signal~\cite{reddy2021dnsmos}.

\begin{figure}[tbp]
\centering
\includegraphics[width=0.65\linewidth]{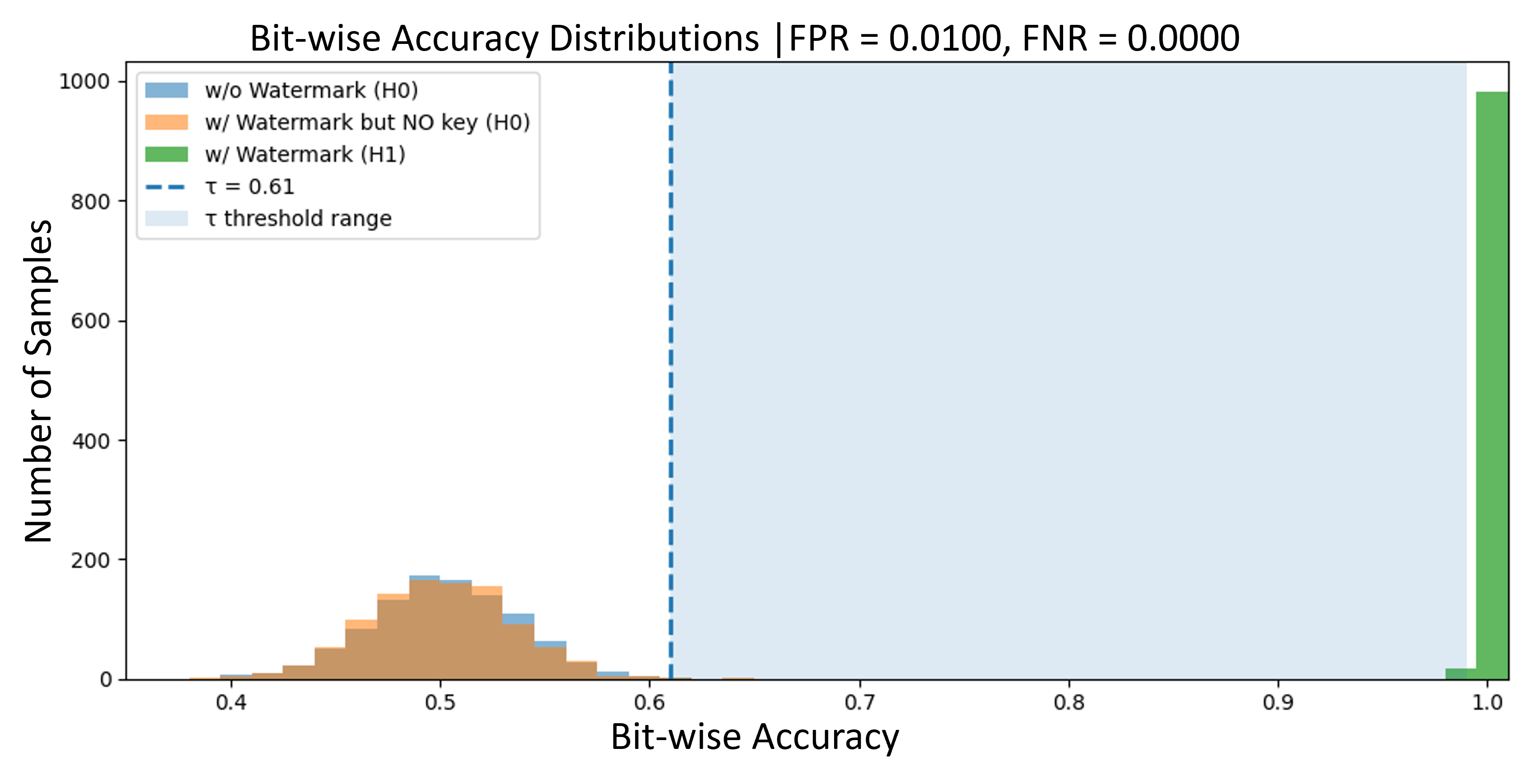}
\caption{Bit-wise accuracy distributions of the key-conditioned verification statistic $S=\mathrm{BitAcc}(\mathbf{m},\hat{\mathbf{m}})$ under $H_0$ (no watermark or wrong key) and $H_1$ (correct key).}
\label{fig:fpr-threshold}
\vspace{-0.1in}
\end{figure}

\subsection{Bit Accuracy}

\paragraph{Key-conditioned verification and thresholding.}
We determine watermark presence using a key-conditioned verification statistic based on bit accuracy. Given a claimed key $K$, the verifier decodes a candidate payload $\hat{\mathbf{m}}$ and computes $S=\mathrm{BitAcc}(\mathbf{m},\hat{\mathbf{m}})$ as defined in Eq.~\eqref{eq:bitacc_def}, where $\mathbf{m}$ denotes the expected payload associated with the claimed user or utterance. Fig.~\ref{fig:fpr-threshold} illustrates the empirical distributions of $S$ over $1000$ samples per condition. Under the alternative hypothesis $H_1$ (watermarked audio evaluated with the correct key), $S$ concentrates near $1.00$, indicating reliable watermark recovery. In contrast, under the null hypothesis $H_0$, including both unwatermarked audio and watermarked audio evaluated with an incorrect key, $S$ concentrates around $0.5$ and exhibits an approximately Gaussian-shaped distribution, consistent with random guessing.

This clear statistical separation enables threshold-based verification. In particular, selecting $\tau = 0.61$ achieves a low false positive rate while maintaining a near-zero false negative rate in our experiments. More generally, any threshold $\tau > 0.61$ lies within a low-density region of the $H_0$ distribution, thereby providing strong statistical evidence of a valid watermark under the claimed key.

\vspace{-0.1in}
\paragraph{Key-based separation.}
Because the spreading patterns are deterministically derived from the secret key, a mismatched key regenerates patterns that are statistically uncorrelated with the embedded watermark residual. Consequently, the resulting verification statistic is indistinguishable from that of the no-watermark case. Therefore, without the correct key, an adversary cannot reliably recover the target payload or exceed bit accuracy threshold for successful verification. 

\begin{figure}[tbp]
\centering
\begin{minipage}[t]{0.49\textwidth}
  \centering
  \includegraphics[width=\linewidth]{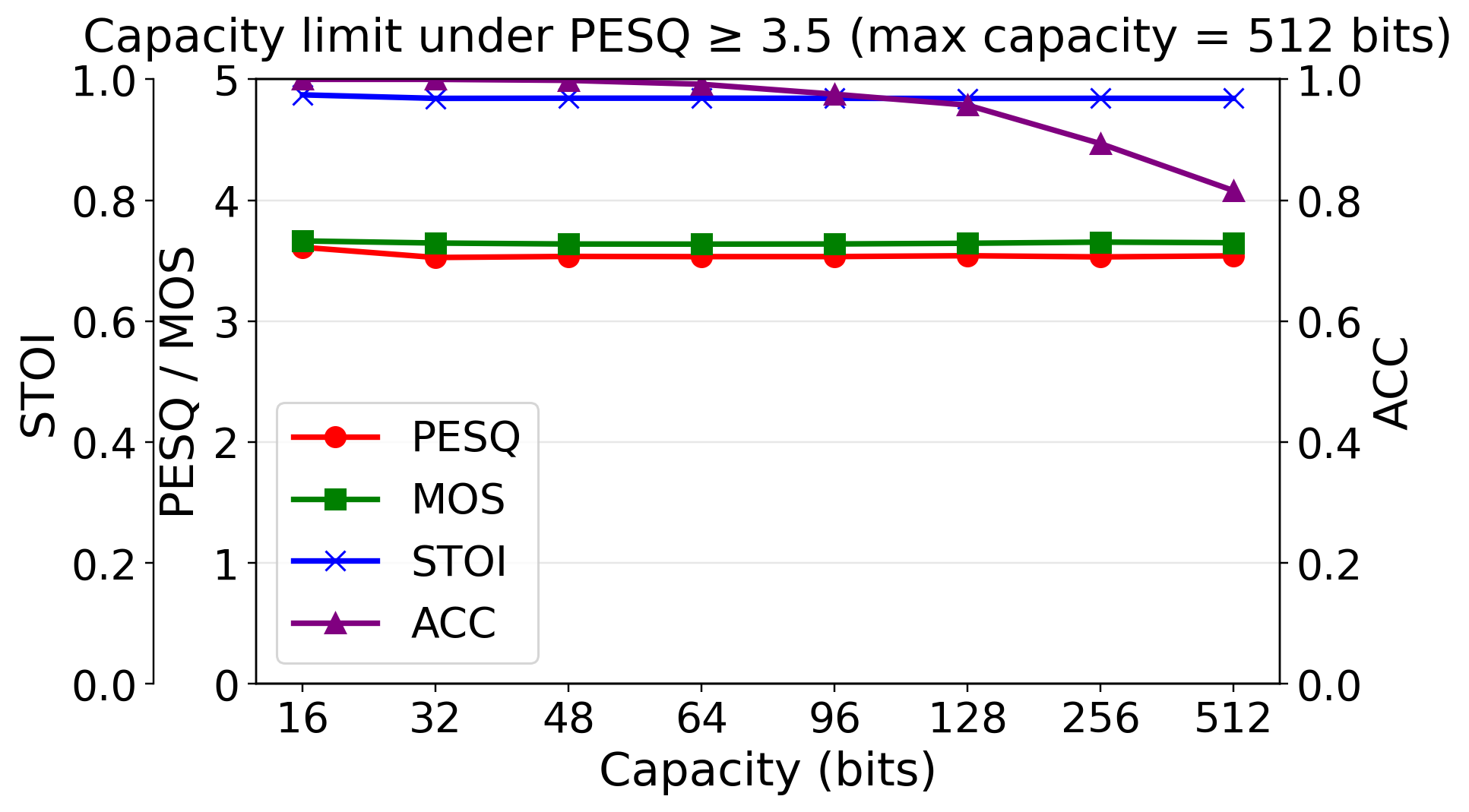}
  \caption{DiffWave: Maximum watermark payload capacity under a perceptual quality constraint (PESQ $\ge$ 3.5).}
  \label{fig:diffwave_cap_quality}
\end{minipage}\hfill
\begin{minipage}[t]{0.49\textwidth}
  \centering
  \includegraphics[width=\linewidth]{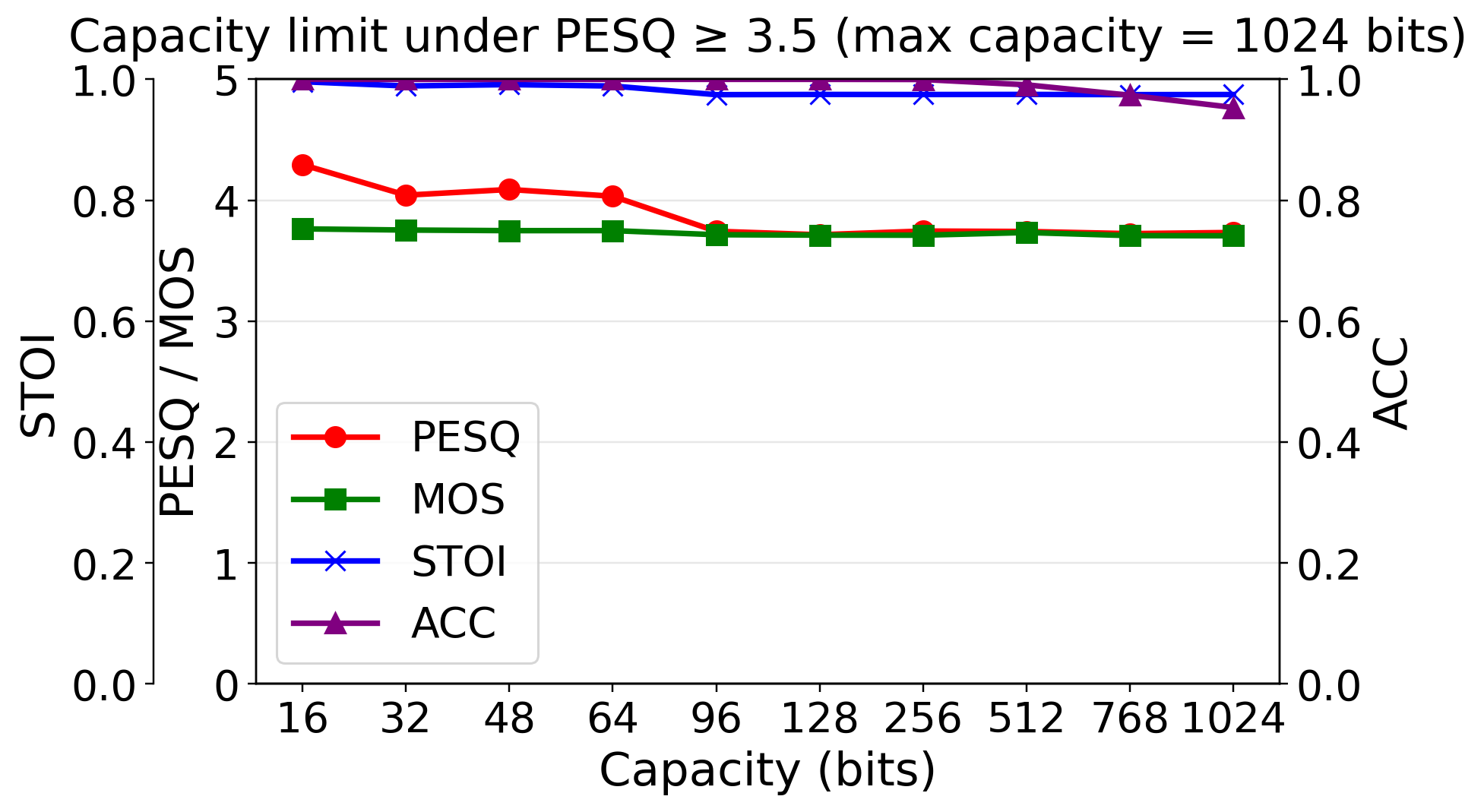}
  \caption{HiFi-GAN: Maximum watermark payload capacity under a perceptual quality constraint (PESQ $\ge$ 3.5).}
  \label{fig:hifigan_cap_quality}
\end{minipage}
\vspace{-0.1in}
\end{figure}

\subsection{Fidelity and Capacity}
\paragraph{Fidelity Performance.}
Table~\ref{tab:fidelity_capacity_combined} shows that MelShield introduces only minor perceptual degradation for both vocoders. To distinguish overall synthesis quality from watermark-induced effects, we report two PESQ variants. PESQ computed with the ground-truth recording as reference reflects end-to-end generation quality, whereas PESQ computed with the unwatermarked synthesis as reference isolates the quality impact introduced by watermark embedding.
For HiFi-GAN, PESQ with the ground-truth reference remains high, exceeding $3.59$ even at a payload of $512$ bits. Meanwhile, STOI remains close to $0.96$ and DNSMOS stays around $3.65$, indicating that speech intelligibility and perceptual naturalness are largely preserved. DiffWave exhibits a similar trend at moderate payload capacities: at $128$ bits, PESQ (ground-truth reference) remains above $3.50$, STOI exceeds $0.96$, and decoding accuracy stays above $0.97$.
Across payload sizes, watermark recovery remains reliable. HiFi-GAN achieves near-perfect decoding performance, with $\mathrm{BitAcc}=1.00$ up to $256$ bits and $0.9965$ at $512$ bits. These results demonstrate that MelShield can scale payload capacity while maintaining a small perceptual quality gap between watermarked and unwatermarked synthesis, particularly for HiFi-GAN operating in high-fidelity regimes.

\vspace{-0.1in}
\paragraph{Payload Capacity Limits Under Quality Constraints.}

Figs.~\ref{fig:diffwave_cap_quality} and~\ref{fig:hifigan_cap_quality} quantify the achievable payload capacity under a high-fidelity operating constraint. Specifically, we enforce a PESQ requirement of at least $3.5$ and evaluate decoding reliability as payload size increases. MelShield sustains strong watermark recovery while scaling to large payload capacities.
For DiffWave, decoding accuracy remains above $95\%$ up to $128$ bits, while PESQ, STOI, and DNSMOS remain stable across the capacity sweep. In contrast, HiFi-GAN supports substantially higher payloads under the same quality requirement: decoding accuracy exceeds $95\%$ even at $1024$ bits, with STOI remaining close to $0.98$ throughout.
These results demonstrate that the Mel-domain spread-spectrum design enables high-capacity watermarking without compromising perceptual quality. Moreover, the achievable capacity depends on the vocoder’s tolerance to log-Mel perturbations, with HiFi-GAN exhibiting notably greater robustness and supporting higher payloads under identical quality constraints.

\begin{table}[t]
\centering
\caption{{Robustness and audio quality under single attacks on DiffWave. We report PESQ ($\uparrow$), STOI ($\uparrow$), MOS ($\uparrow$), and decoding accuracy ACC ($\uparrow$).}
\label{tab:robust-diffwave}}
\setlength{\tabcolsep}{3.5pt}
\small
\resizebox{\columnwidth}{!}{%
\begin{tabular}{llcccccccccc}
\toprule
Method (DiffWave) & Metric & MP3 & AAC & Sca & Rs16 & BP & LP & N20 & N10 & N5 & Echo \\
\midrule
\multirow{4}{*}{AudioSeal} & PESQ & 4.4413 & 4.4153 & 4.4418 & 4.4417 & 4.4226 & 4.4035 & 1.4610 & 1.0707 & 1.0331 & 1.4344 \\
& STOI & 0.9986 & 0.9985 & 0.9986 & 0.9986 & 0.9924 & 0.9983 & 0.9764 & 0.9138 & 0.8569 & 0.9481 \\
& MOS & 3.6572 & 3.6533 & 3.6804 & 3.6484 & 3.6560 & 3.5644 & 2.9916 & 2.5791 & 2.3928 & 3.1054 \\
& ACC & 0.9925 & 0.8113 & 0.9981 & 0.9994 & 0.9981 & 0.9738 & 0.9537 & 0.7644 & 0.6150 & 0.9988 \\
\midrule
\multirow{4}{*}{WavMark} & PESQ & 3.8537 & 4.0968 & 4.1457 & 4.1436 & 1.4929 & 3.7341 & 1.3942 & 1.0575 & 1.0304 & 1.4398 \\
& STOI & 0.9984 & 0.9982 & 0.9984 & 0.9984 & 0.9923 & 0.9980 & 0.9725 & 0.8996 & 0.8383 & 0.9486 \\
& MOS & 3.6227 & 3.6359 & 3.6562 & 3.6346 & 3.5487 & 3.5494 & 2.9707 & 2.5427 & 2.3532 & 3.1305 \\
& ACC & 0.8719 & 0.8750 & 0.8925 & 0.8688 & 0.8644 & 0.8644 & 0.6025 & 0.5188 & 0.4944 & 0.8506 \\
\midrule
\multirow{4}{*}{Timbre} & PESQ & 3.6395 & 3.6086 & 3.6428 & 3.6566 & 3.6682 & 3.7783 & 1.4443 & 1.0689 & 1.0327 & 1.4086 \\
& STOI & 0.9955 & 0.9954 & 0.9955 & 0.9955 & 0.9894 & 0.9953 & 0.9748 & 0.9090 & 0.8509 & 0.9470 \\
& MOS & 3.6462 & 3.6407 & 3.6726 & 3.6416 & 3.6276 & 3.5852 & 2.9797 & 2.5741 & 2.3827 & 3.0854 \\
& ACC & \textbf{1.0000} & \textbf{1.0000} & \textbf{1.0000} & \textbf{1.0000} & \textbf{1.0000} & \textbf{1.0000} & {0.9517} & 0.7125 & 0.5725 & \textbf{1.0000} \\
\midrule
\multirow{4}{*}{GROOT} & PESQ & 3.1372 & 3.1098 & 3.3191 & 3.2708 & 3.0543 & 3.3385 & 1.4041 & 0.9976 & 0.8714 & 1.1136 \\
& STOI & 0.9301 & 0.9279 & 0.9290 & 0.9130 & 0.8113 & 0.9339 & 0.9055 & 0.8726 & 0.8010 & 0.8961 \\
& MOS  & 3.3703 & 3.3858 & 3.3991 & 3.3779 & 3.3108 & 3.3692 & 2.8870 & 2.3431 & 2.1904 & 2.9502 \\
& ACC  & 0.9959 & 0.9945 & \textbf{1.0000} & \textbf{1.0000} & \textbf{1.0000} & \textbf{1.0000} & \textbf{0.9901} & \textbf{0.9878} & \textbf{0.9421} & 0.9834 \\
\midrule
\multirow{4}{*}{MelShield} & PESQ & 3.4077 & 3.3945 & 3.4089 & 3.4079 & 3.4747 & 3.4194 & 1.4380 & 1.0723 & 1.0334 & 1.3824 \\
& STOI & 0.9602 & 0.9601 & 0.9602 & 0.9602 & 0.9553 & 0.9600 & 0.9424 & 0.8913 & 0.8414 & 0.9133 \\
& MOS & 3.6170 & 3.6097 & 3.6499 & 3.6163 & 3.5952 & 3.5546 & 2.9342 & 2.5550 & 2.3595 & 3.0473 \\
& ACC & \textbf{1.0000} & \textbf{1.0000} & \textbf{1.0000} & \textbf{1.0000} & \textbf{1.0000} & \textbf{1.0000} & 0.9519 & 0.7788 & 0.7006 & \textbf{1.0000} \\
\bottomrule
\end{tabular}%
}
\vspace{-8pt}
\end{table}

\begin{table}[t]
\centering
\caption{{Robustness and audio quality under single attacks on HiFi-GAN. We report PESQ ($\uparrow$), STOI ($\uparrow$), MOS ($\uparrow$), and decoding accuracy ACC ($\uparrow$).}
\label{tab:robust-hifigan}}
\setlength{\tabcolsep}{3.5pt}
\small
\resizebox{\columnwidth}{!}{%
\begin{tabular}{llcccccccccc}
\toprule
Method (HiFi-GAN) & Metric
& MP3 & AAC & Sca
& Rs16
& BP & LP
& N20 & N10 & N5
& Echo \\
\midrule
\multirow{4}{*}{AudioSeal} & PESQ & 4.4321 & 4.4017 & 4.4389 & 4.4375 & 4.4108 & 4.3962 & 1.4723 & 1.0819 & 1.0384 & 1.4289 \\
& STOI & 0.9985 & 0.9984 & 0.9985 & 0.9985 & 0.9920 & 0.9982 & 0.9758 & 0.9146 & 0.8582 & 0.9469 \\
& MOS  & 3.6528 & 3.6469 & 3.6761 & 3.6427 & 3.6493 & 3.5582 & 2.9824 & 2.5716 & 2.3871 & 3.0978 \\
& ACC  & 0.9876 & 0.8049 & 0.9969 & 0.9989 & 0.9968 & 0.9692 & 0.9501 & 0.7587 & 0.6064 & 0.9979 \\
\midrule
\multirow{4}{*}{WavMark} & PESQ & 3.8612 & 4.0885 & 4.1519 & 4.1397 & 1.5016 & 3.7289 & 1.4018 & 1.0612 & 1.0311 & 1.4362 \\
& STOI & 0.9983 & 0.9981 & 0.9983 & 0.9983 & 0.9921 & 0.9979 & 0.9719 & 0.9007 & 0.8396 & 0.9479 \\
& MOS  & 3.6189 & 3.6322 & 3.6517 & 3.6295 & 3.5421 & 3.5448 & 2.9642 & 2.5368 & 2.3479 & 3.1237 \\
& ACC  & 0.8684 & 0.8728 & 0.8891 & 0.8652 & 0.8609 & 0.8617 & 0.5963 & 0.5126 & 0.4879 & 0.8468 \\
\midrule
\multirow{4}{*}{Timbre} & PESQ & 3.6458 & 3.6149 & 3.6504 & 3.6632 & 3.6725 & 3.7856 & 1.4527 & 1.0744 & 1.0346 & 1.4121 \\
& STOI & 0.9954 & 0.9953 & 0.9954 & 0.9954 & 0.9891 & 0.9952 & 0.9743 & 0.9098 & 0.8520 & 0.9462 \\
& MOS  & 3.6410 & 3.6358 & 3.6679 & 3.6371 & 3.6220 & 3.5809 & 2.9728 & 2.5669 & 2.3759 & 3.0796 \\
& ACC & \textbf{1.0000} & \textbf{1.0000} & \textbf{1.0000} & \textbf{1.0000} & \textbf{1.0000} & 0.9987 & 0.9463 & 0.7062 & 0.5654 & 0.9996 \\
\midrule
\multirow{4}{*}{MelShield} & PESQ & 4.1342 & 4.1415 & 4.1498 & 4.1137 & 4.1802 & 4.1268 & 1.4456 & 1.0759 & 1.0342 & 1.3897 \\
& STOI & 0.9605 & 0.9603 & 0.9604 & 0.9605 & 0.9557 & 0.9602 & 0.9430 & 0.8921 & 0.8423 & 0.9145 \\
& MOS  & 3.6127 & 3.6062 & 3.6461 & 3.6119 & 3.5906 & 3.5508 & 2.9289 & 2.5498 & 2.3557 & 3.0414 \\
& ACC  & \textbf{1.0000} & \textbf{1.0000} & \textbf{1.0000} & \textbf{1.0000} & \textbf{1.0000} & \textbf{0.9996} & \textbf{1.0000} & \textbf{0.7815} & \textbf{0.7052} & \textbf{1.0000} \\
\bottomrule
\end{tabular}%
}
\vspace{-14pt}
\end{table}

\begin{figure}[htbp]
\centering

\subfloat[16 bits\label{fig:alpha16}]{%
  \includegraphics[width=0.44\textwidth]{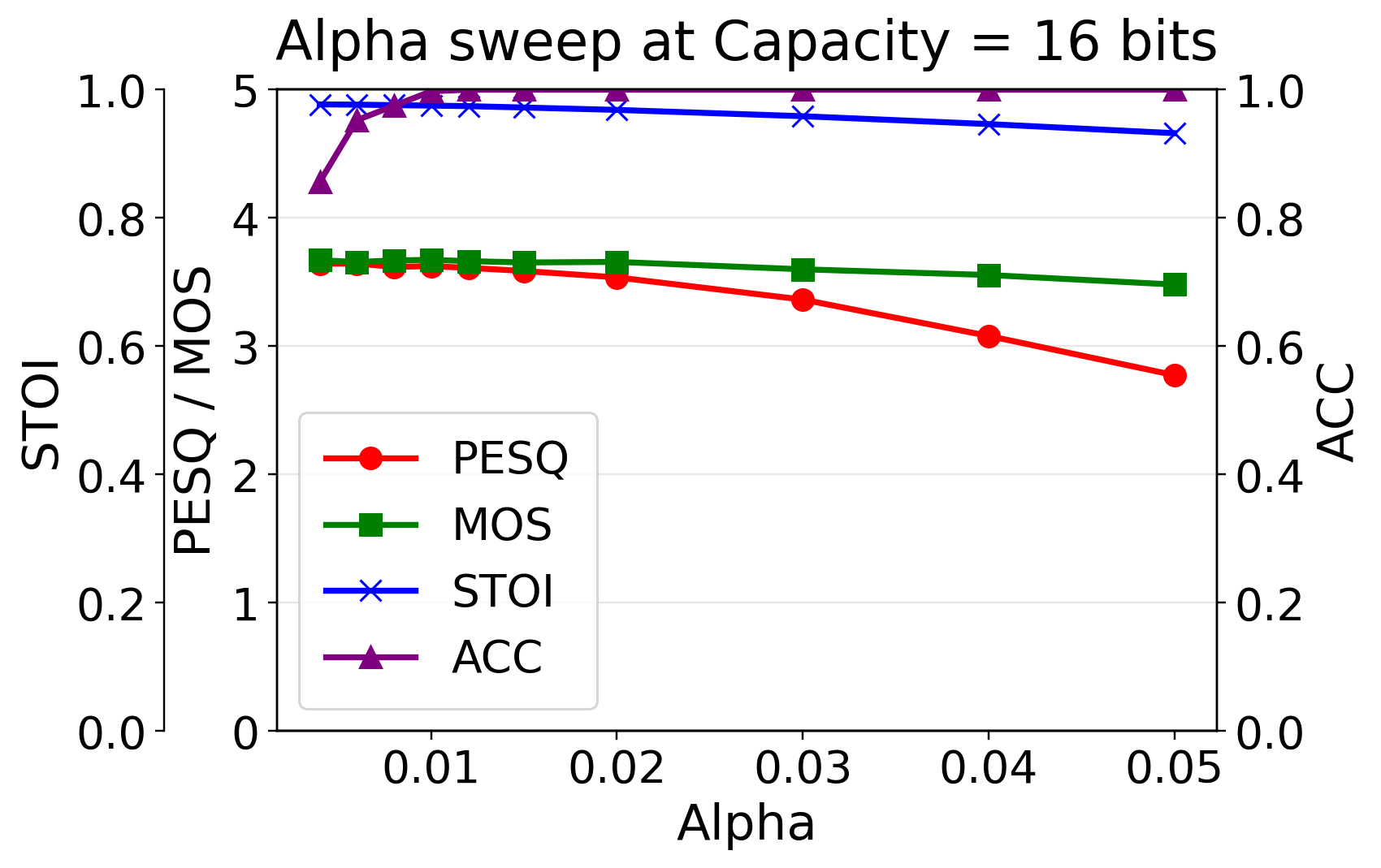}
}\hfill
\subfloat[32 bits\label{fig:alpha32}]{%
  \includegraphics[width=0.44\textwidth]{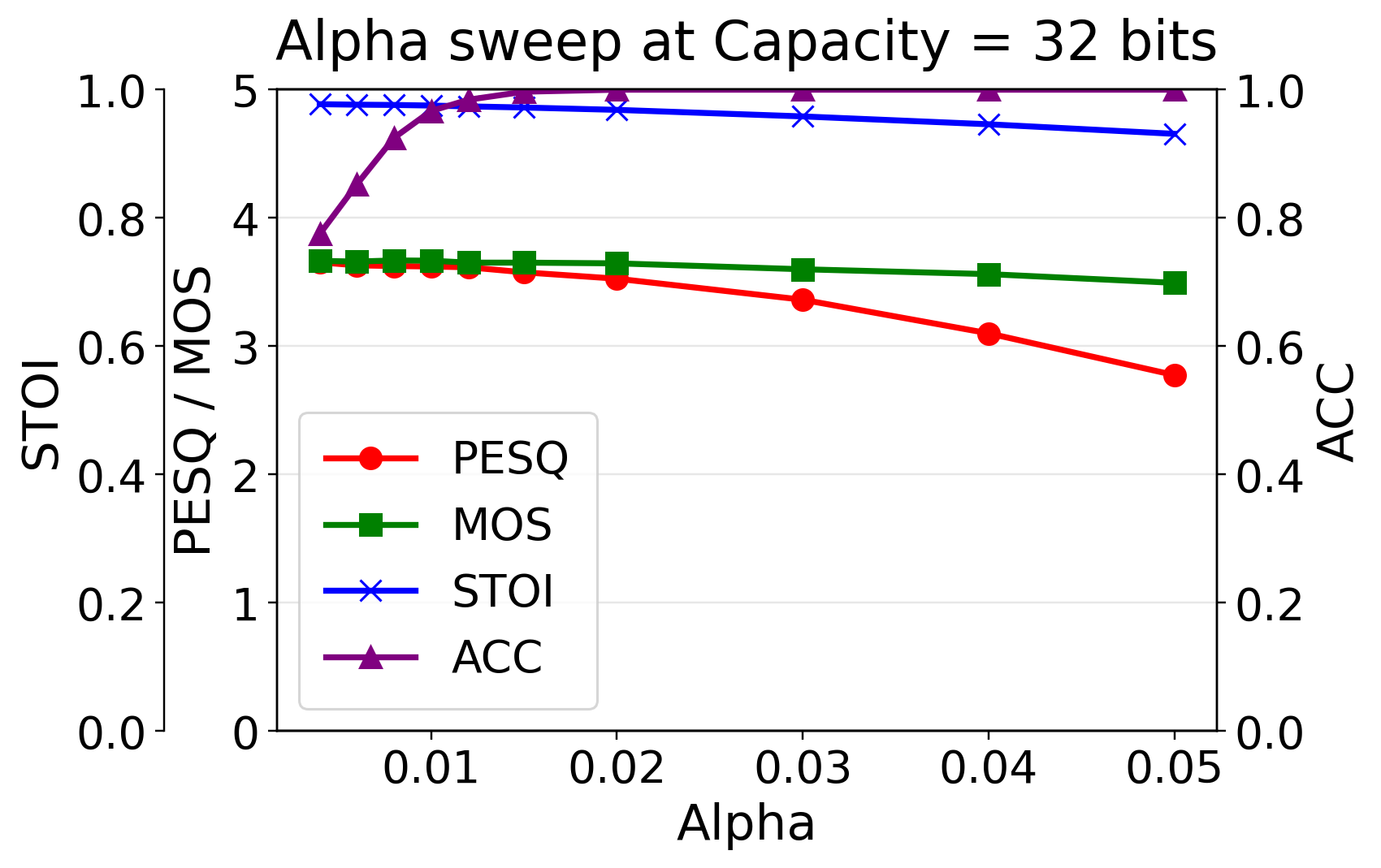}
}

\vspace{1mm}

\subfloat[48 bits\label{fig:alpha48}]{%
  \includegraphics[width=0.44\textwidth]{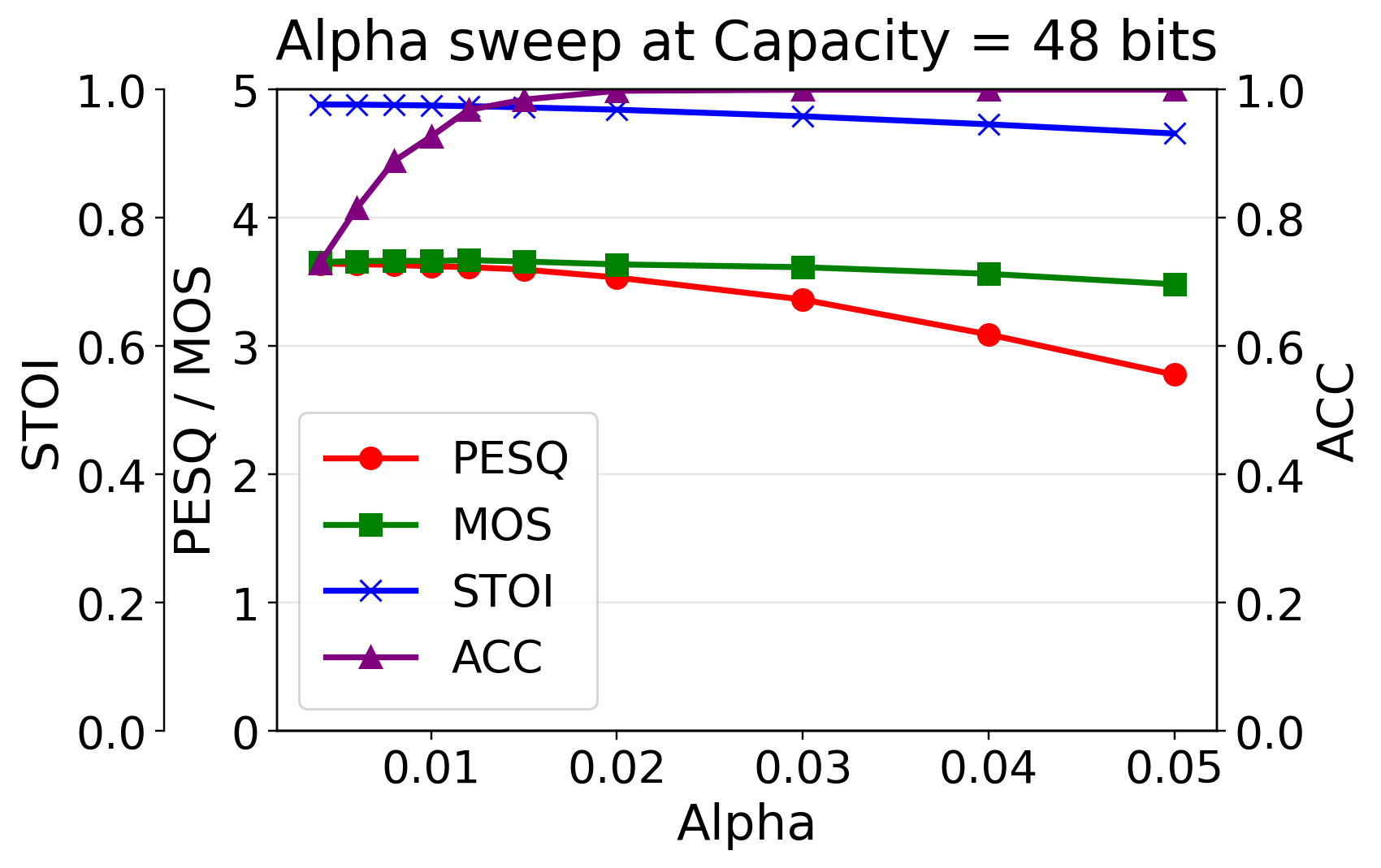}
}\hfill
\subfloat[64 bits\label{fig:alpha64}]{%
  \includegraphics[width=0.44\textwidth]{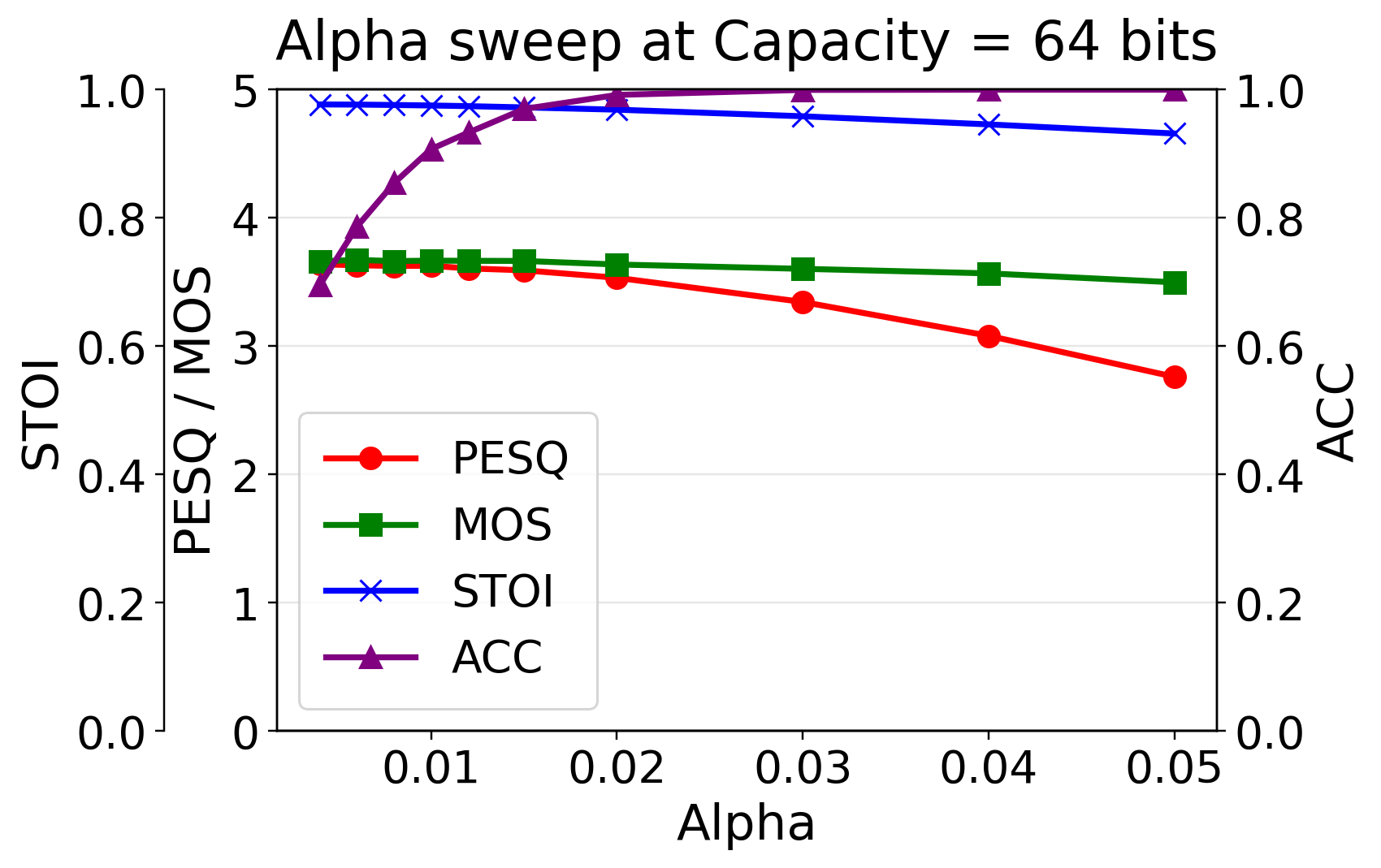}
}

\vspace{1mm}

\subfloat[96 bits\label{fig:alpha96}]{%
  \includegraphics[width=0.44\textwidth]{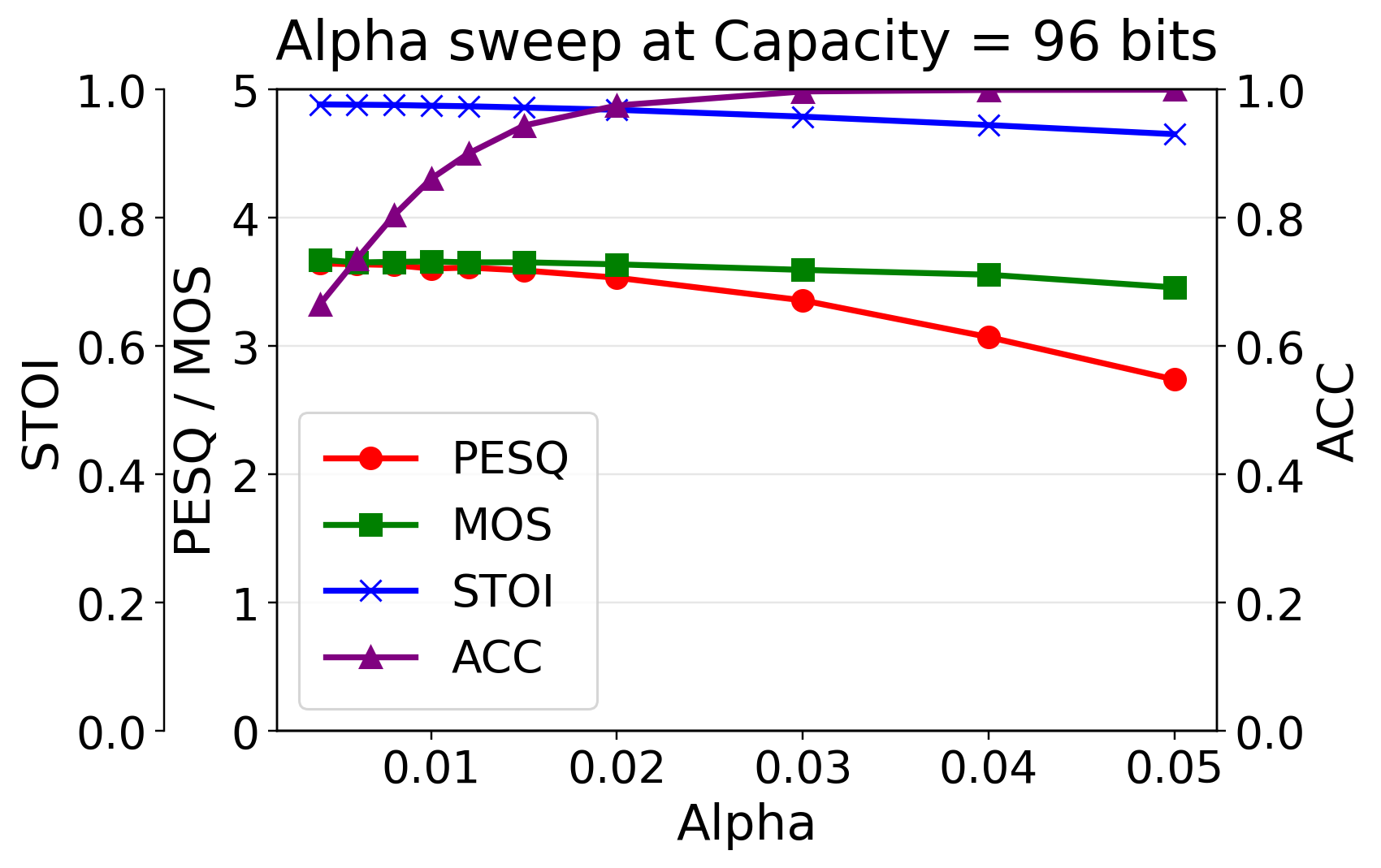}
}\hfill
\subfloat[128 bits\label{fig:alpha128}]{%
  \includegraphics[width=0.44\textwidth]{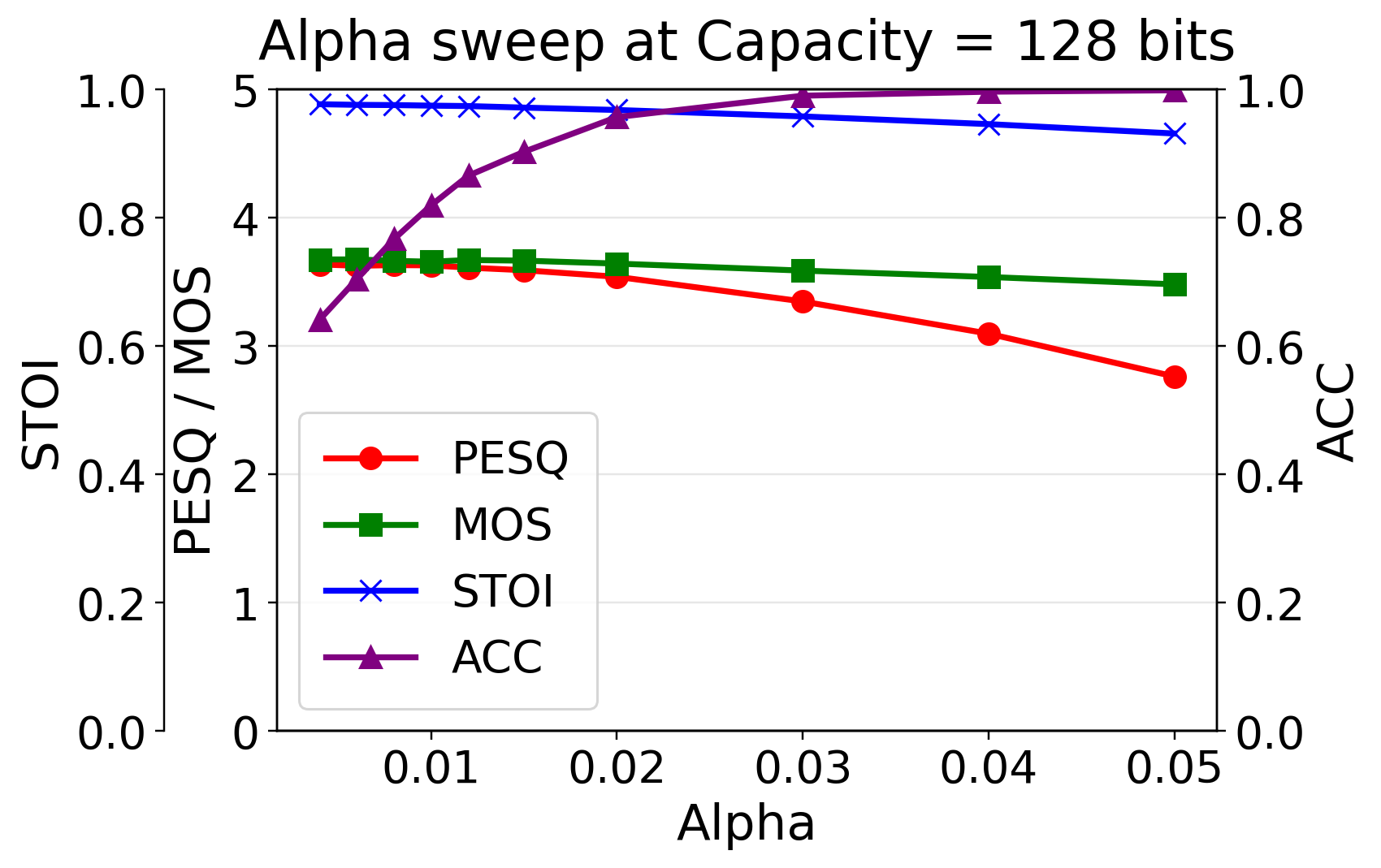}
}

\vspace{1mm}

\subfloat[256 bits\label{fig:alpha256}]{%
  \includegraphics[width=0.44\textwidth]{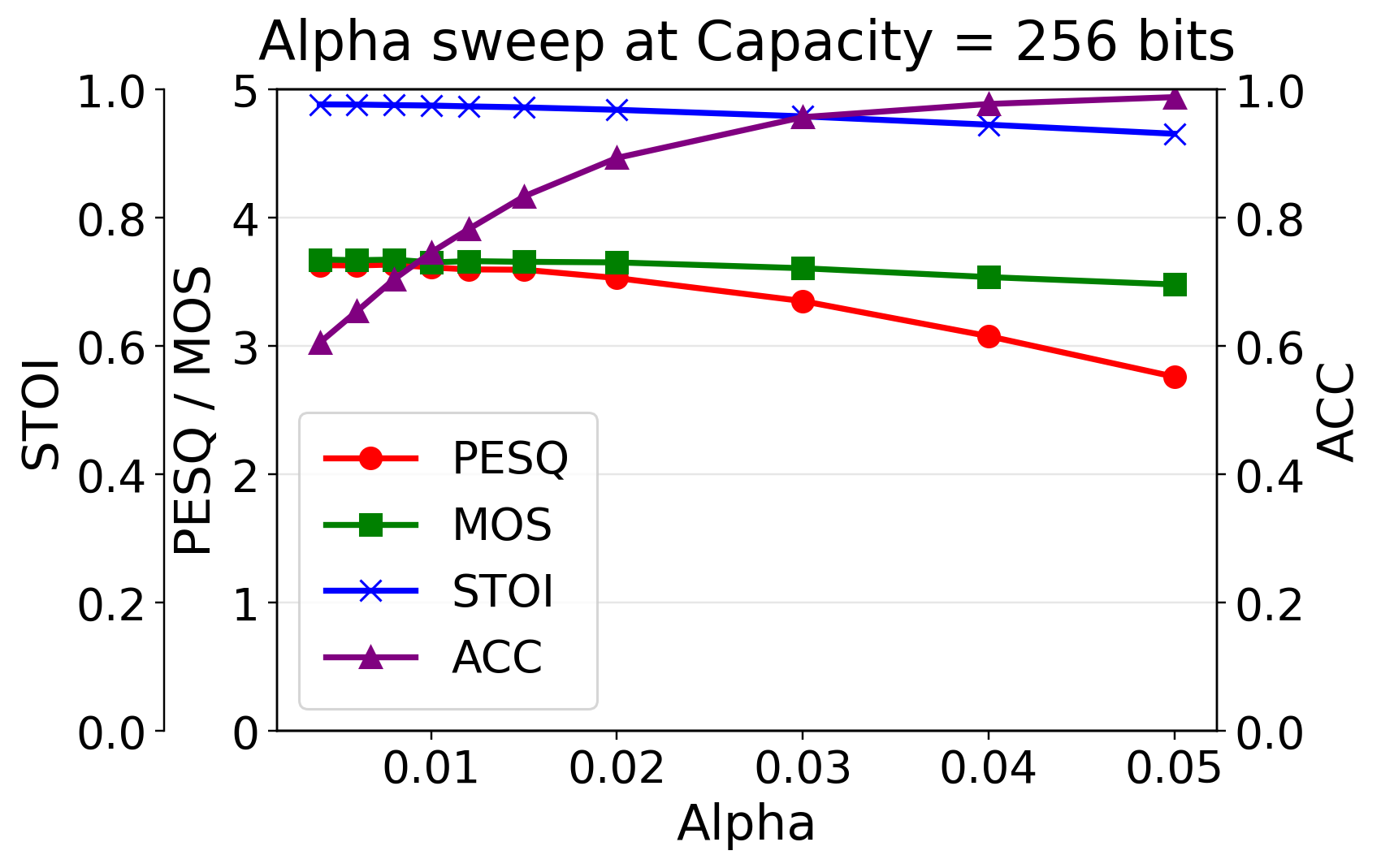}
}\hfill
\subfloat[512 bits\label{fig:alpha512}]{%
  \includegraphics[width=0.44\textwidth]{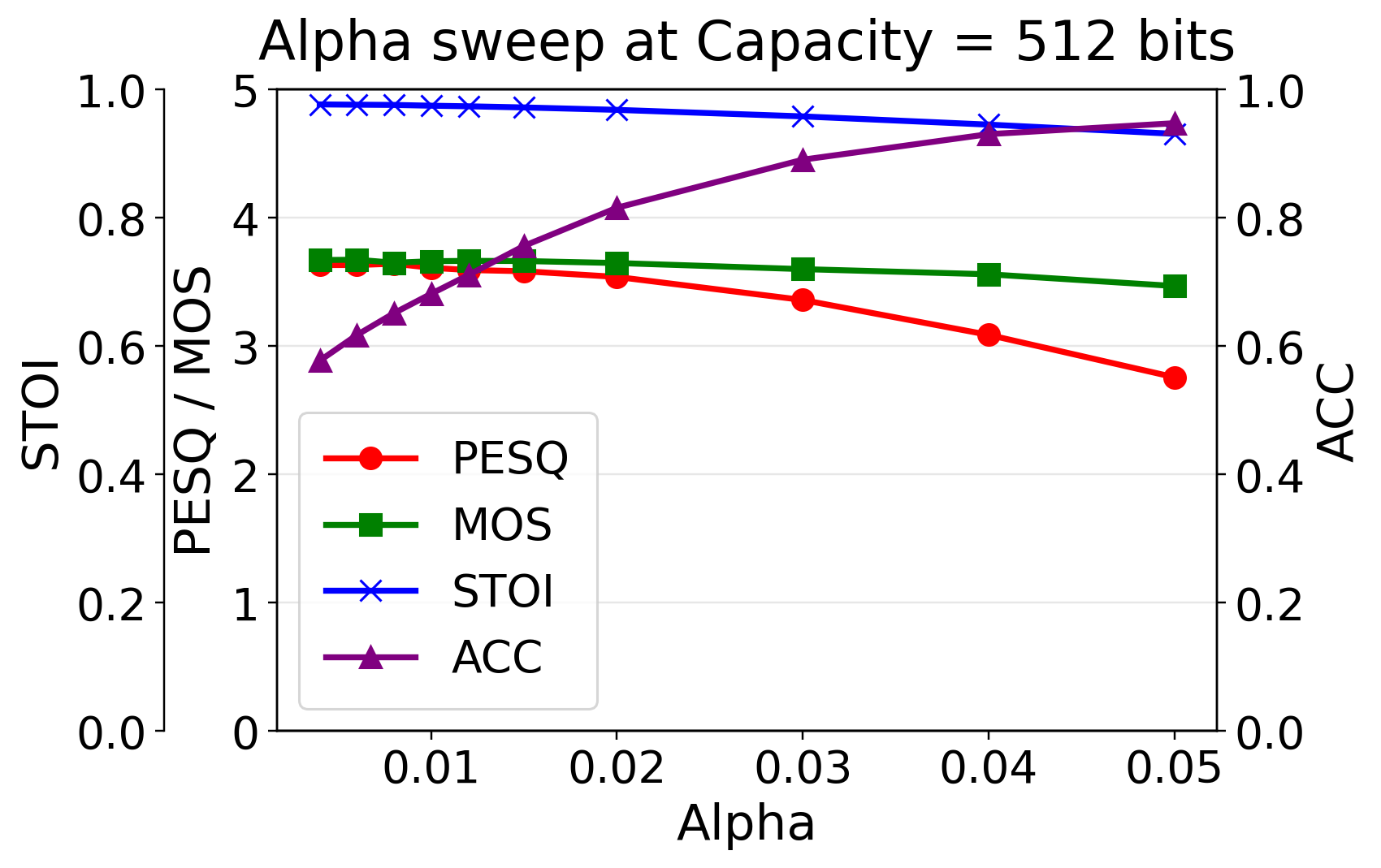}
}

\caption{DiffWave: Alpha sweep under different payload capacities. Each subfigure shows PESQ, MOS, STOI, and decoding accuracy as a function of the embedding strength $\alpha$.}
\label{fig:alpha_sweep_2x4}
\end{figure}

\begin{figure}[htbp]
\centering

\subfloat[16 bits\label{fig:hifi_alpha16}]{\includegraphics[width=0.4\textwidth]{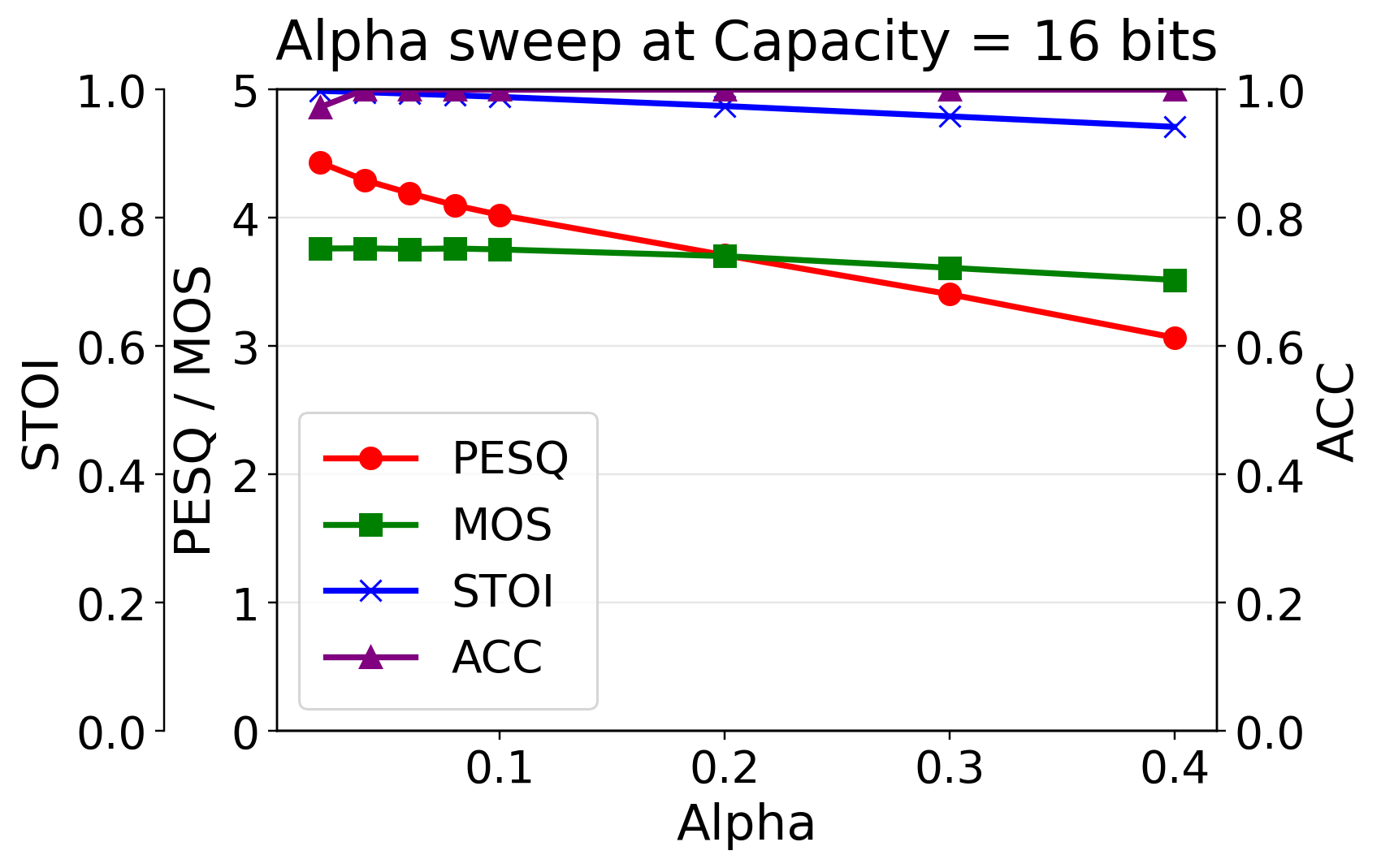}}\hfill
\subfloat[32 bits\label{fig:hifi_alpha32}]{\includegraphics[width=0.4\textwidth]{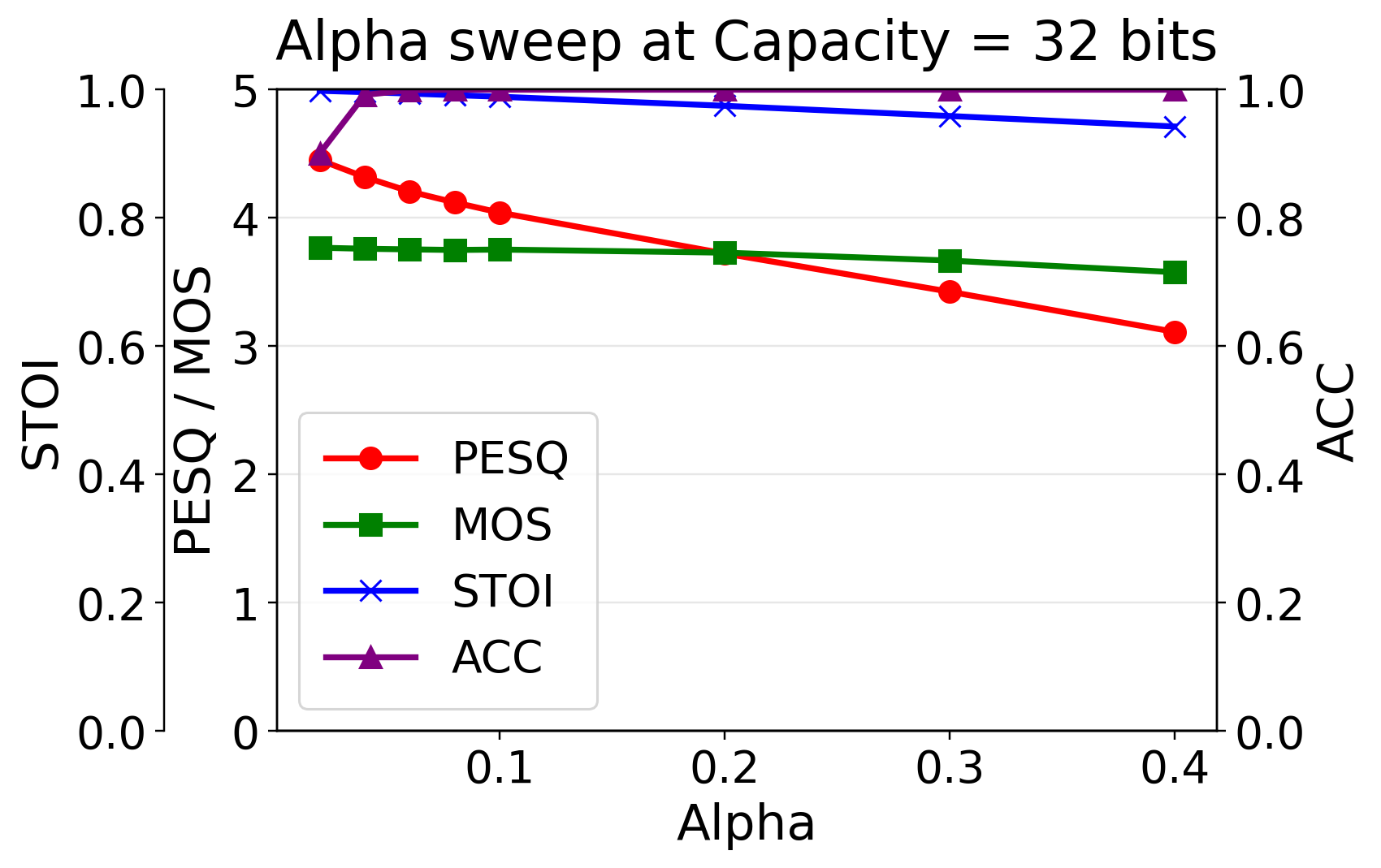}}


\subfloat[48 bits\label{fig:hifi_alpha48}]{\includegraphics[width=0.4\textwidth]{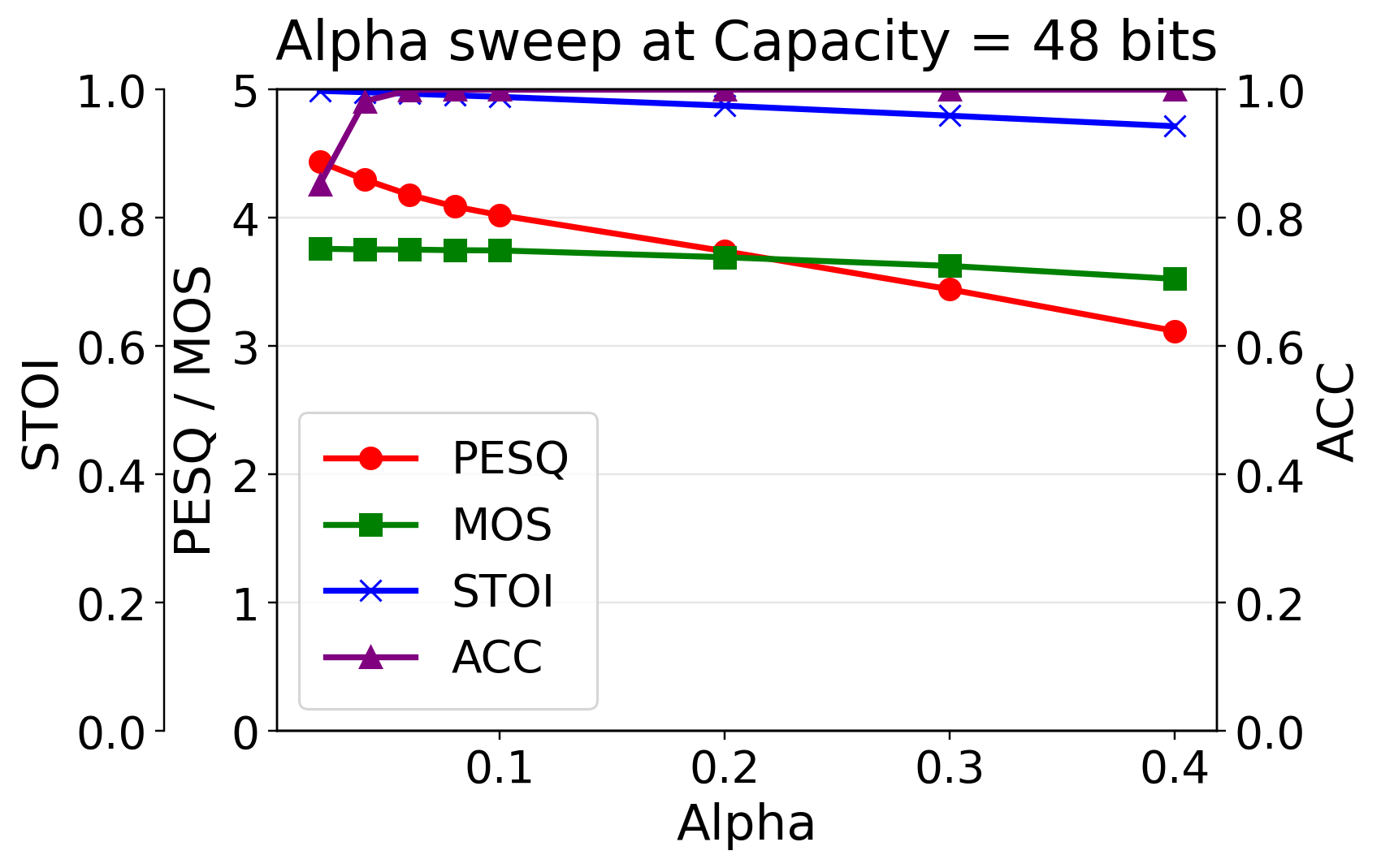}}\hfill
\subfloat[64 bits\label{fig:hifi_alpha64}]{\includegraphics[width=0.4\textwidth]{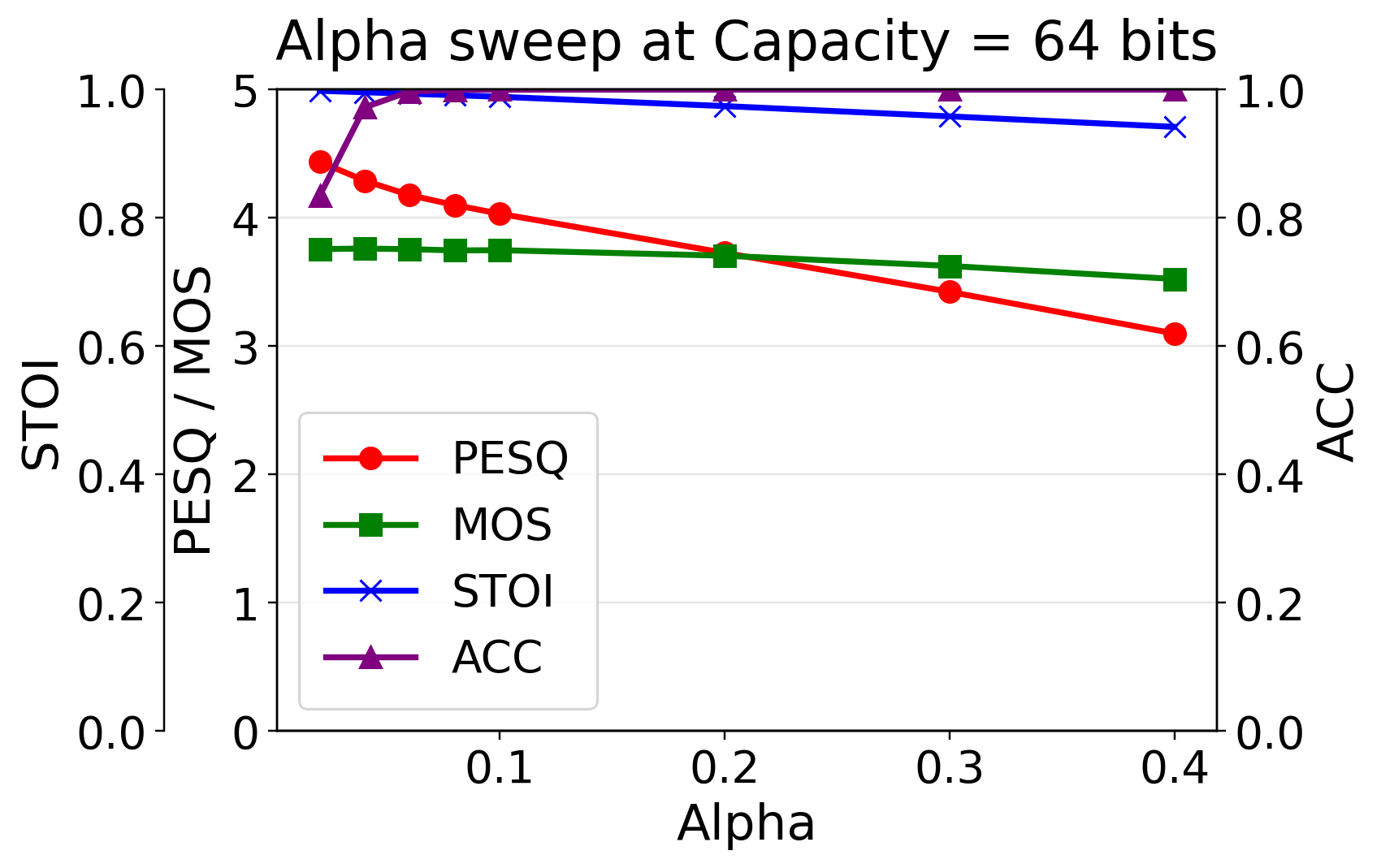}}


\subfloat[96 bits\label{fig:hifi_alpha96}]{\includegraphics[width=0.4\textwidth]{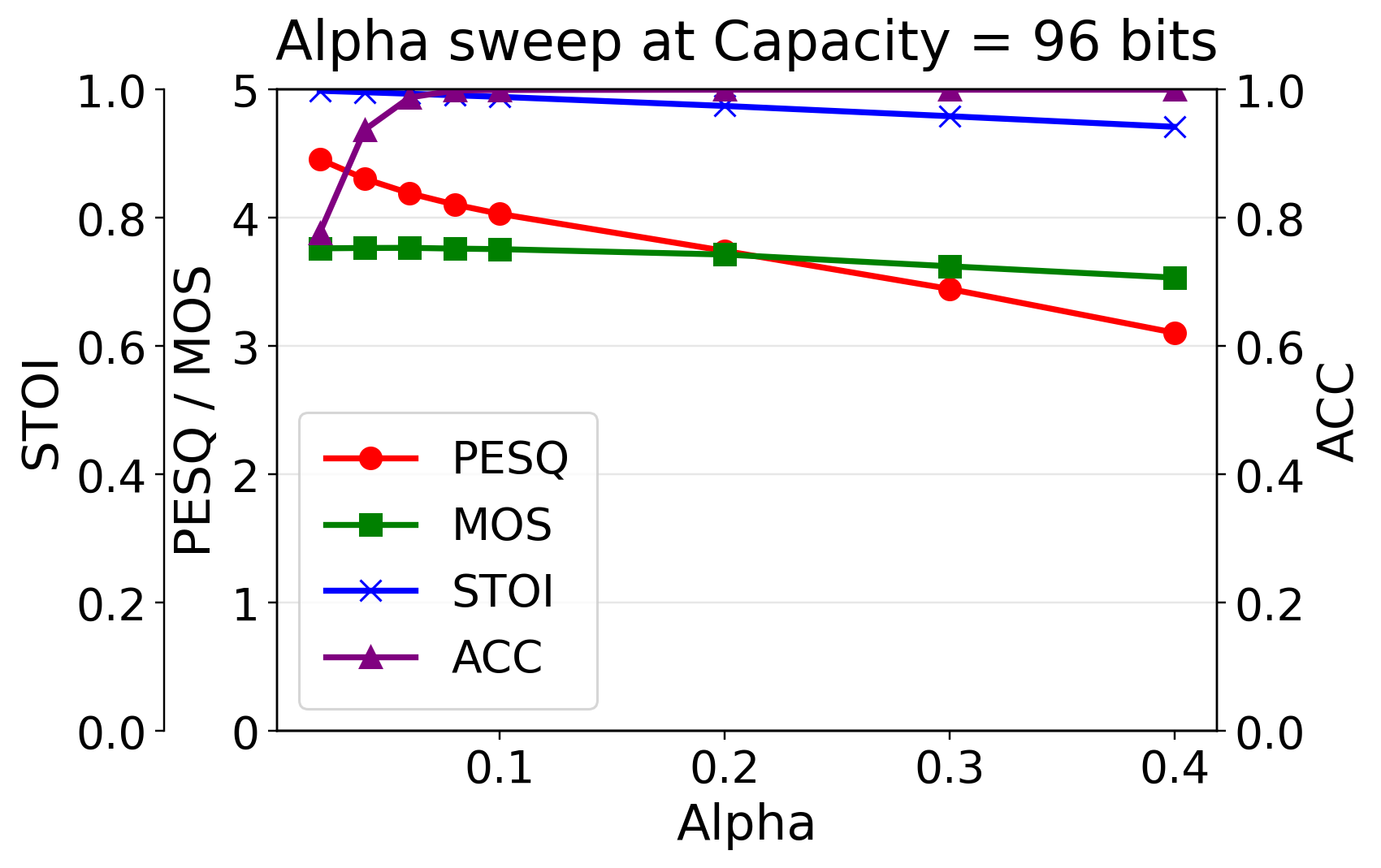}}\hfill
\subfloat[128 bits\label{fig:hifi_alpha128}]{\includegraphics[width=0.4\textwidth]{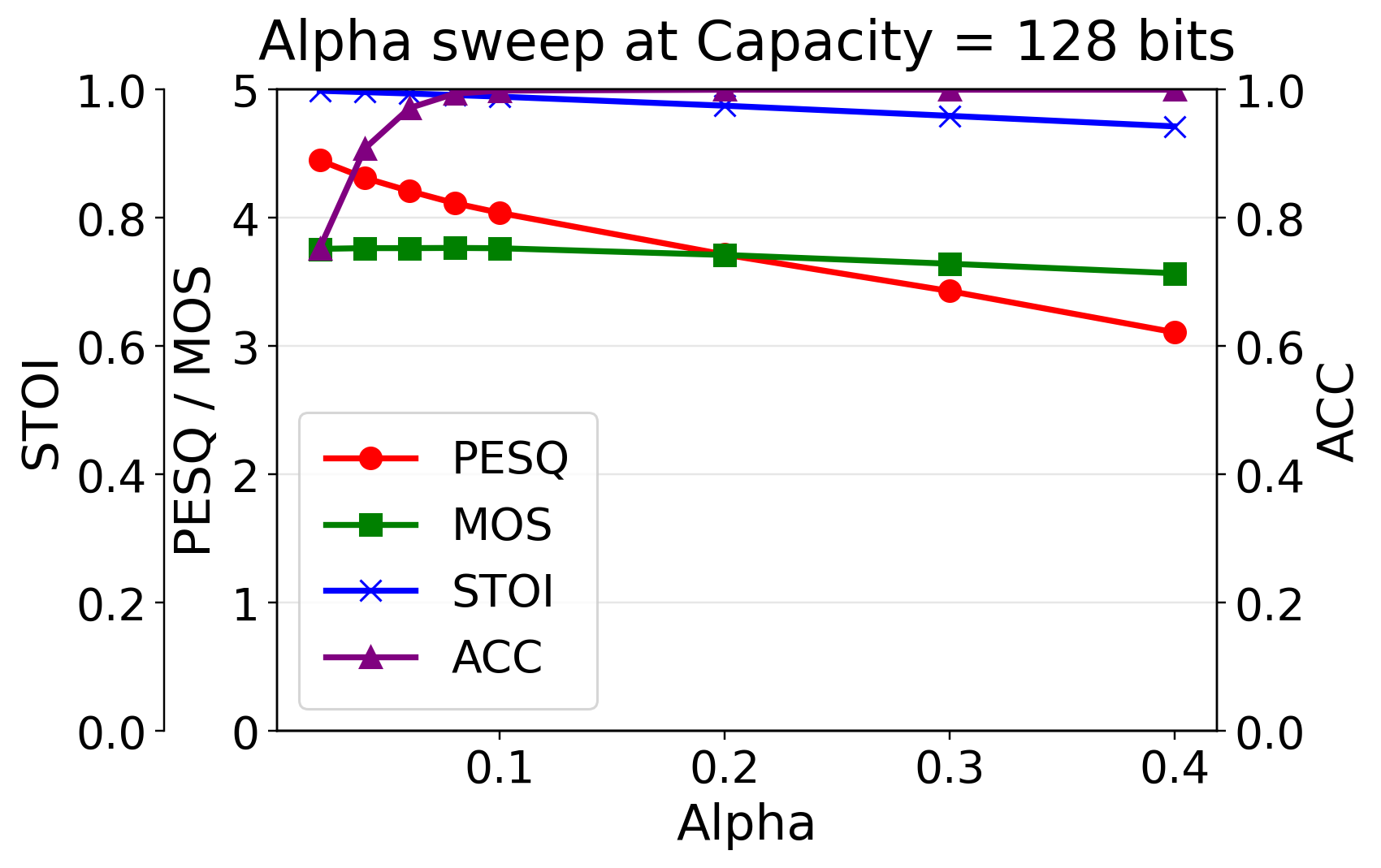}}


\subfloat[256 bits\label{fig:hifi_alpha256}]{\includegraphics[width=0.4\textwidth]{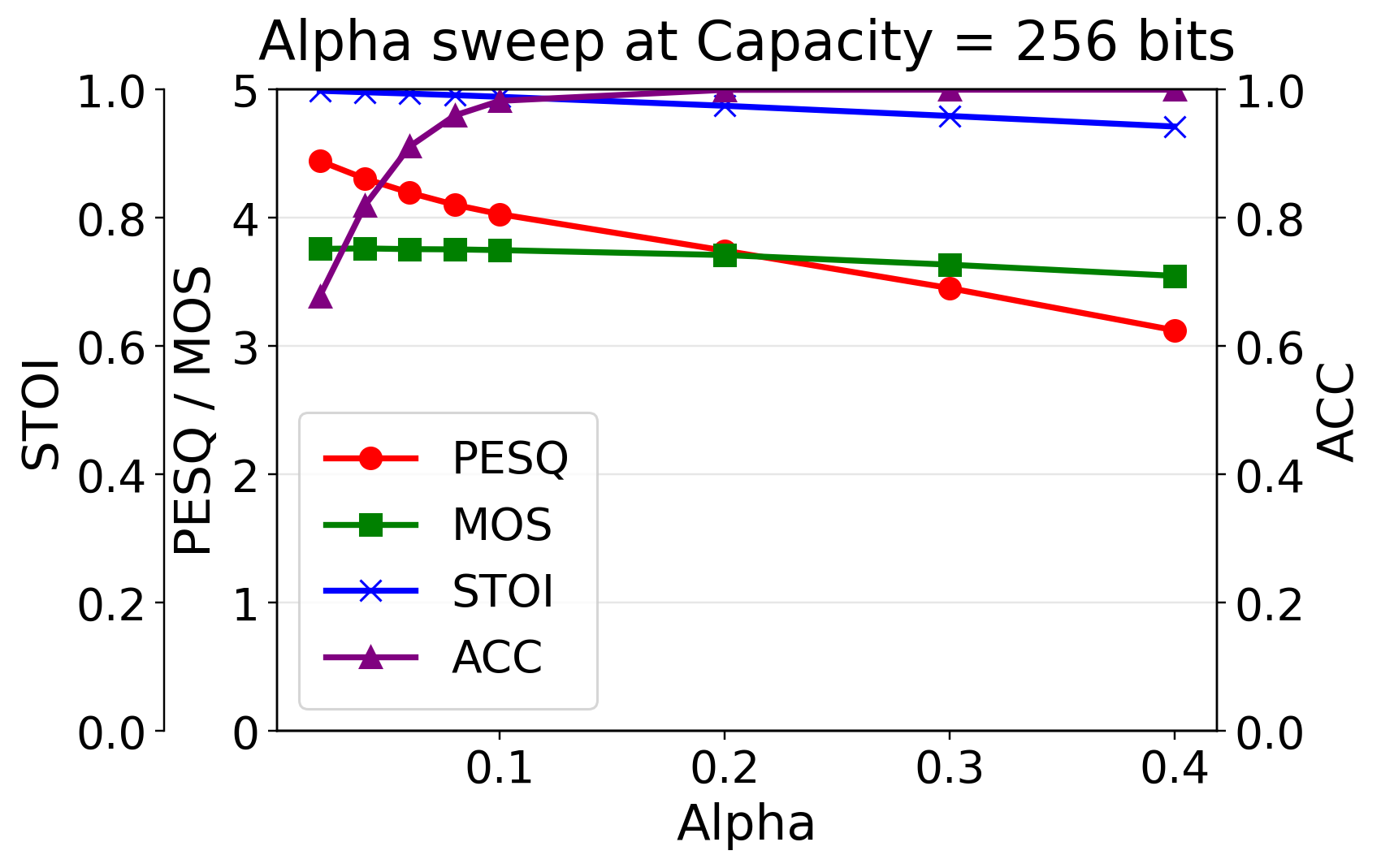}}\hfill
\subfloat[512 bits\label{fig:hifi_alpha512}]{\includegraphics[width=0.4\textwidth]{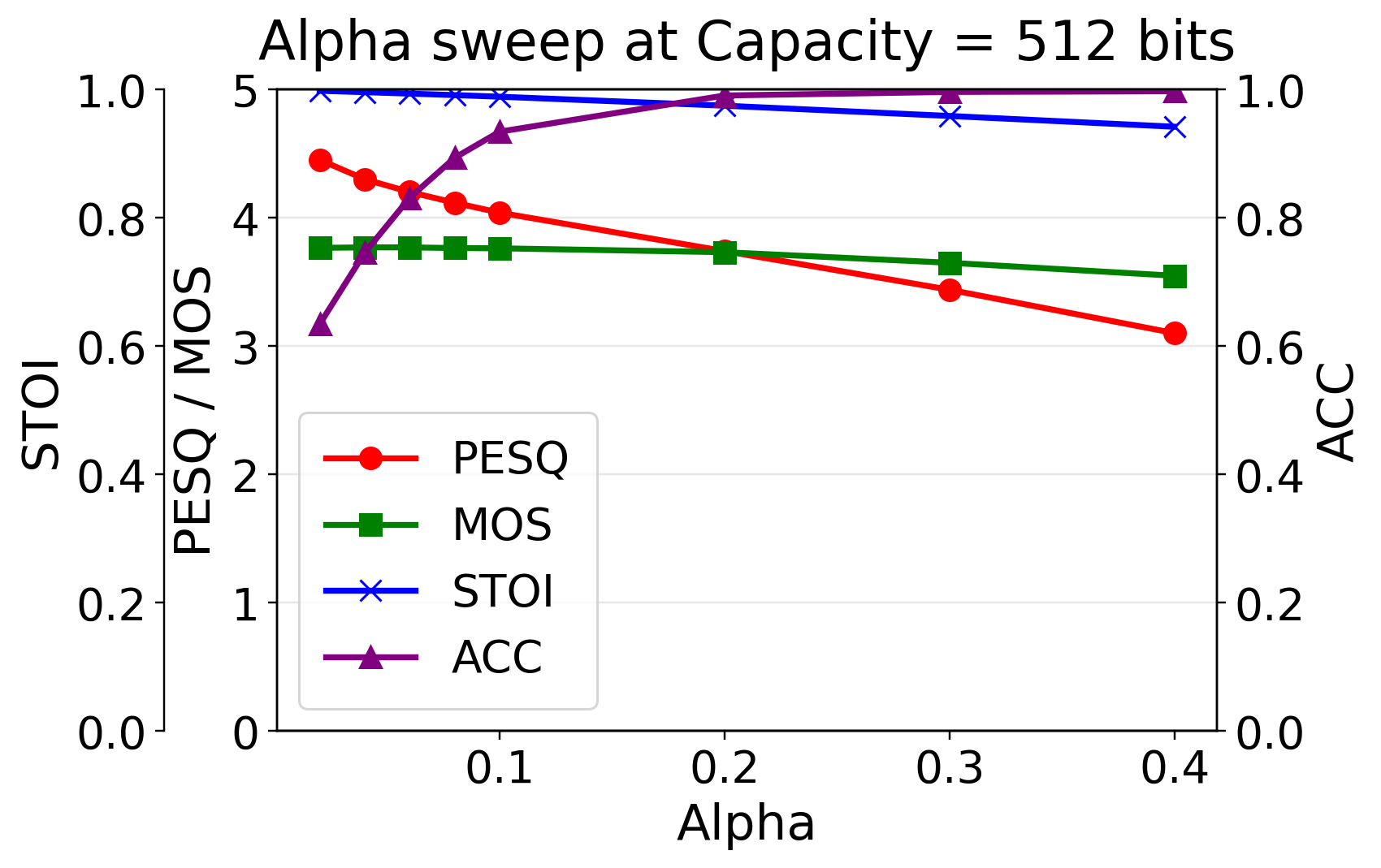}}


\subfloat[768 bits\label{fig:hifi_alpha768}]{\includegraphics[width=0.4\textwidth]{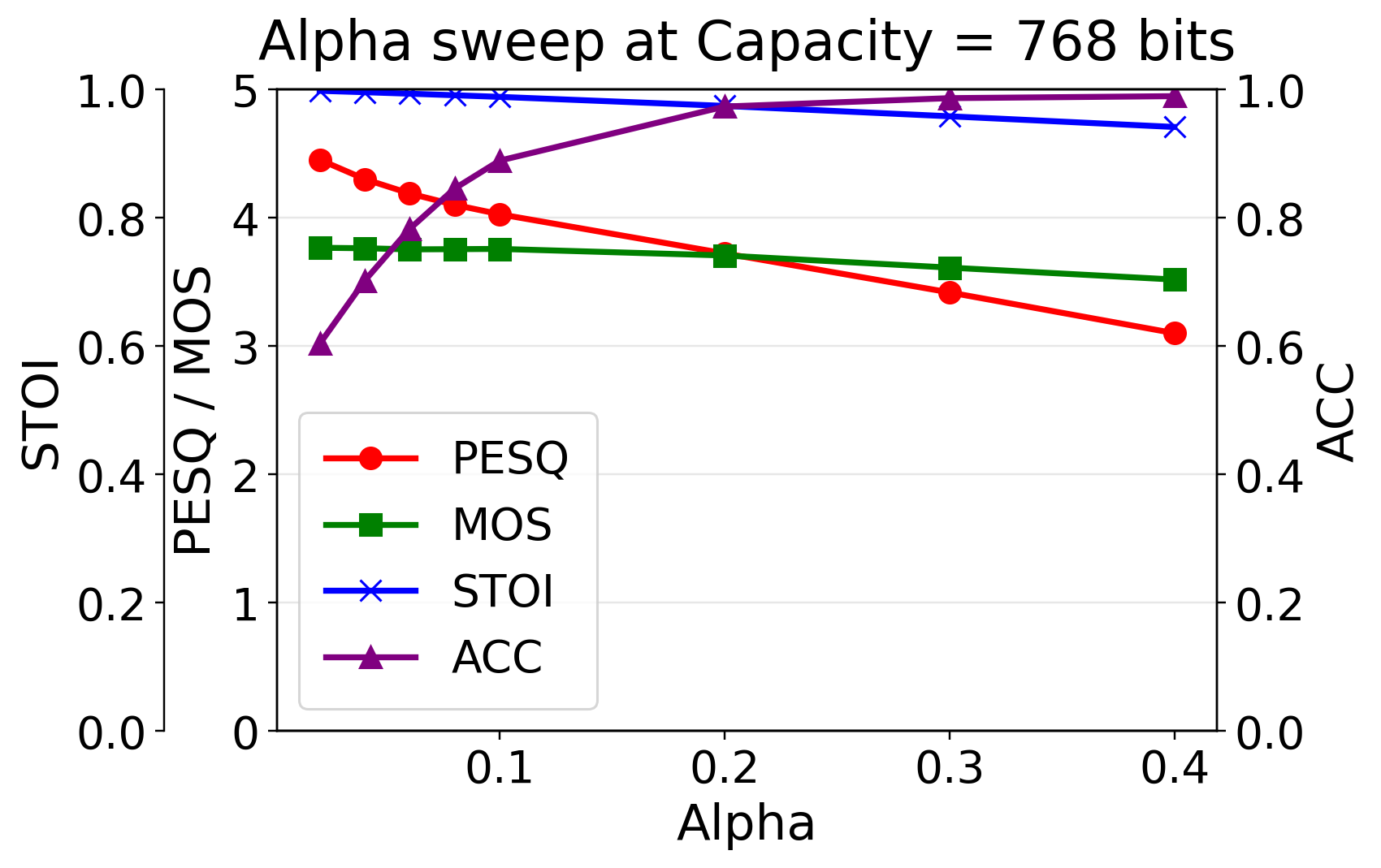}}\hfill
\subfloat[1024 bits\label{fig:hifi_alpha1024}]{\includegraphics[width=0.4\textwidth]{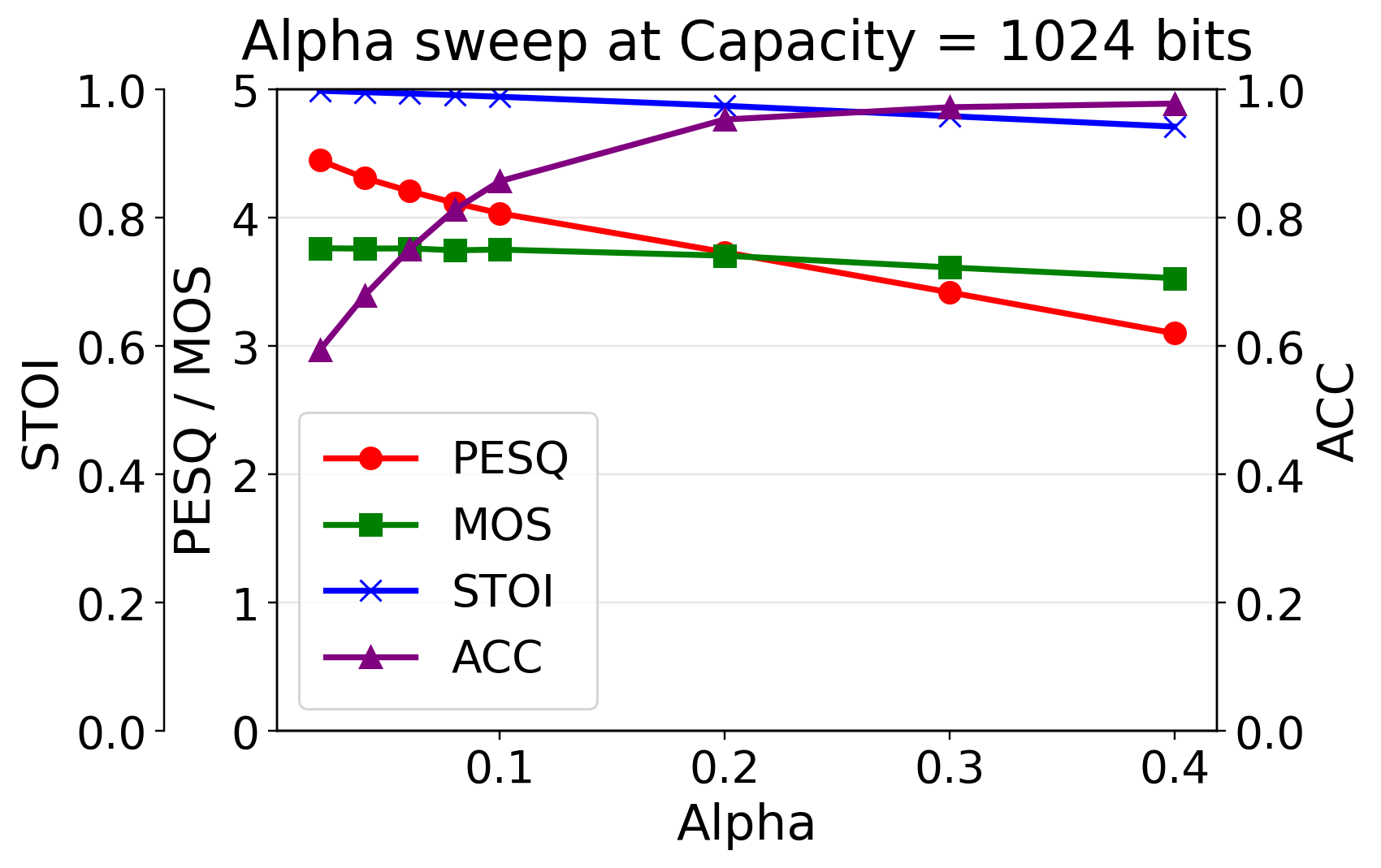}}

\caption{HiFi-GAN: Alpha sweep under different payload capacities. Each subfigure reports PESQ, MOS, STOI, and decoding accuracy as a function of the embedding strength $\alpha$.}
\label{fig:hifigan_alpha_sweep_2x5}
\end{figure}

\vspace{-0.1in}
\paragraph{Trade-off between Accuracy, Fidelity, and Capacity.}
Figs.~\ref{fig:alpha_sweep_2x4} and \ref{fig:hifigan_alpha_sweep_2x5} present an $\alpha$ sweep under different payload capacities. A consistent robustness-fidelity trade-off is observed. As $\alpha$ increases, bit-wise decoding accuracy improves monotonically, with the improvement being most pronounced at higher capacities where the detector must distinguish a larger number of superposed bit patterns. However, increasing $\alpha$ also introduces stronger perturbations in the log-Mel domain, leading to gradual degradation in perceptual quality metrics. PESQ exhibits the clearest downward trend as $\alpha$ grows, while DNSMOS shows a similar but more moderate decline. In contrast, STOI remains comparatively stable and stays close to $0.98$ across a wide range of $\alpha$, indicating that speech intelligibility is largely preserved even at higher embedding strengths. At low payload capacities (e.g., $16$--$64$ bits), decoding accuracy saturates rapidly and approaches $1.0$ even at small $\alpha$, suggesting that the system can operate in a near-lossless regime. As the payload increases to $512$ and $1024$ bits, accuracy at small $\alpha$ becomes noticeably lower, requiring larger embedding strengths to achieve high $\mathrm{BitAcc}$. This behavior directly reflects the intrinsic capacity--fidelity trade-off in Mel-domain spread-spectrum watermarking.
\vspace{-0.1in}

\paragraph{Practical selection of $\alpha$.}

A potential strategy is to impose an explicit perceptual quality constraint and then select the smallest $\alpha$ that achieves a desired decoding accuracy. For example, under a PESQ constraint of $3.5$, the curves exhibit a clear elbow region: $\mathrm{BitAcc}$ increases rapidly with only minor perceptual degradation, whereas further increasing $\alpha$ yields diminishing gains in accuracy while continuing to reduce PESQ and DNSMOS.
The location of this elbow depends on the payload capacity. Smaller payloads typically permit lower $\alpha$ values while achieving near-perfect decoding performance. In contrast, larger payloads require moderately higher $\alpha$ to maintain $\mathrm{BitAcc} \ge 0.95$, reflecting the inherent capacity-fidelity trade-off.
When higher perceptual fidelity is prioritized, one may reduce $\alpha$ and compensate by (i) lowering the payload size, (ii) increasing audio duration to enable stronger statistical evidence aggregation during verification, or (iii) selecting a more conservative operating point on the decoding accuracy curve. These design choices provide flexible deployment trade-offs tailored to different attribution requirements.

\subsection{Robustness}

We compare MelShield against three representative post-hoc baselines (WavMark, AudioSeal, and Timbre) under typical signal processing distortions on LJSpeech. For each method, a 32-bit payload is embedded, and we report the average bit-wise decoding accuracy after applying additive noise at $5/10/20$~dB SNR, low-pass filtering at $3$~kHz (LP-F~3k), band-pass filtering between $0.3$--$8$~kHz (BP-F~0.3--8k), amplitude scaling, echo addition, resampling to $16$~kHz, and lossy compression using MP3-128 and AAC-96 with default codec settings.
As summarized in Table~\ref{tab:robust-diffwave} and Table~\ref{tab:robust-hifigan}, MelShield instantiated with both DiffWave and HiFi-GAN vocoders demonstrates strong robustness under additive noise while remaining highly competitive under filtering, resampling, and compression. In particular, decoding accuracy remains above $0.99$ across all non-noise conditions and exceeds $0.95$ at $20$~dB SNR, consistently outperforming the compared baselines. GROOT may exhibit stronger robustness under certain extreme noise settings; however, such gains are typically accompanied by substantially larger perceptual quality degradation under practical operating regimes. Overall, MelShield achieves a more favorable robustness-fidelity trade-off, maintaining strong decoding reliability while preserving audio quality.



\section{Conclusion}
In this paper, we presented MelShield, a model-agnostic in-generation audio watermarking framework that works directly in the Mel domain of Mel-conditioned TTS pipelines. MelShield enables keyed spread-spectrum embedding for user-specific attribution without modifying or retraining the underlying vocoder. Its plug-and-play design makes it compatible with both diffusion-based and GAN-based vocoders, providing a practical pathway for large-scale deployment across diverse neural speech synthesis systems. Unlike post-hoc methods that can be bypassed or removed as an auxiliary processing stage, MelShield integrates watermarking directly into the generation pipeline, strengthening resilience in real-world distribution environments. Meanwhile, its keyed verification mechanism supports scalable multi-user attribution while reducing the risk of extractor abuse and adversarial probing. Extensive evaluations demonstrate that MelShield achieves a favorable robustness–fidelity–capacity trade-off, validating its suitability for practical AI-generated speech provenance protection.

\bibliographystyle{splncs04}
\bibliography{ref}

@article{van2016wavenet,
  title={Wavenet: A generative model for raw audio},
  author={Van Den Oord, Aaron and Dieleman, Sander and Zen, Heiga and Simonyan, Karen and Vinyals, Oriol and Graves, Alex and Kalchbrenner, Nal and Senior, Andrew and Kavukcuoglu, Koray and others},
  journal={arXiv preprint arXiv:1609.03499},
  volume={12},
  pages={1},
  year={2016}
}

@article{stevens1937mel,
  title   = {A Scale for the Measurement of the Psychological Magnitude Pitch},
  author  = {Stevens, S. S. and Volkmann, J. and Newman, E. B.},
  journal = {The Journal of the Acoustical Society of America},
  volume  = {8},
  number  = {3},
  pages   = {185--190},
  year    = {1937}
}

@misc{zhou2024traceablespeech,
  title={TraceableSpeech: Towards Proactively Traceable Text-to-Speech with Watermarking},
  author={Junzuo Zhou and Jiangyan Yi and Tao Wang and Jianhua Tao and Ye Bai and Chu Yuan Zhang and Yong Ren and Zhengqi Wen},
  url={https://arxiv.org/abs/2406.04840},
  year={2024}
}

@misc{ito2017ljspeech,
  author       = {Keith Ito and Linda Johnson},
  title        = {The LJ Speech Dataset},
  year         = {2017},
  howpublished = {\url{https://keithito.com/LJ-Speech-Dataset/}}
}

@misc{liu2025xattnmark,
  title={XAttnMark: Learning Robust Audio Watermarking with Cross-Attention},
  author={Yixin Liu and Lie Lu and Jihui Jin and Lichao Sun and Andrea Fanelli},
  year={2025},
  eprint={2502.04230},
  archivePrefix={arXiv},
  primaryClass={cs.SD},
  url={https://arxiv.org/abs/2502.04230}
}

@inproceedings{kim2020glowtts,
author = {Kim, Jaehyeon and Kim, Sungwon and Kong, Jungil and Yoon, Sungroh},
title = {Glow-TTS: a generative flow for text-to-speech via monotonic alignment search},
year = {2020},
booktitle = {Proc. of NeurIPS},
articleno = {676},
numpages = {11}
}

@inproceedings{lee2022bigvgan,
  title={BigVGAN: A Universal Neural Vocoder with Large-Scale Training},
  author={Lee, Sang-gil and Ping, Wei and Ginsburg, Boris and Catanzaro, Bryan and Yoon, Sungroh},
  booktitle={Proc. of International Conference on Learning Representations},
  year={2023}
}

@misc{wang2023valle,
  title={Neural Codec Language Models are Zero-Shot Text to Speech Synthesizers},
  author={Chengyi Wang and Sanyuan Chen and Yu Wu and Ziqiang Zhang and Long Zhou and Shujie Liu and Zhuo Chen and Yanqing Liu and Huaming Wang and Jinyu Li and Lei He and Sheng Zhao and Furu Wei},
  year={2023},
  eprint={2301.02111},
  archivePrefix={arXiv},
  primaryClass={cs.CL},
  doi={10.48550/arXiv.2301.02111}
}

@article{chen2023wavmark,
  title={Wavmark: Watermarking for audio generation},
  author={Chen, Guangyu and Wu, Yu and Liu, Shujie and Liu, Tao and Du, Xiaoyong and Wei, Furu},
  journal={arXiv preprint arXiv:2308.12770},
  year={2023}
}

@article{roman2024proactive,
  title={Proactive detection of voice cloning with localized watermarking},
  author={Roman, Robin San and Fernandez, Pierre and D{\'e}fossez, Alexandre and Furon, Teddy and Tran, Tuan and Elsahar, Hady},
  journal={arXiv preprint arXiv:2401.17264},
  year={2024}
}

@inproceedings{liu2023detecting,
  title = {Detecting Voice Cloning Attacks via Timbre Watermarking},
  author = {Liu, Chang and Zhang, Jie and Zhang, Tianwei and Yang, Xi and Zhang, Weiming and Yu, Nenghai},
  booktitle = {Network and Distributed System Security Symposium},
  year = {2024}
}

@inproceedings{
shen2018natural,
  title={Natural tts synthesis by conditioning wavenet on mel spectrogram predictions},
  author={Shen, Jonathan and Pang, Ruoming and Weiss, Ron J and Schuster, Mike and Jaitly, Navdeep and Yang, Zongheng and Chen, Zhifeng and Zhang, Yu and Wang, Yuxuan and Skerrv-Ryan, Rj and others},
  booktitle={Proc. of ICASSP},
  pages={4779--4783},
  year={2018}
}

@article{ren2020fastspeech,
  title={Fastspeech 2: Fast and high-quality end-to-end text to speech},
  author={Ren, Yi and Hu, Chenxu and Tan, Xu and Qin, Tao and Zhao, Sheng and Zhao, Zhou and Liu, Tie-Yan},
  journal={arXiv preprint arXiv:2006.04558},
  year={2020}
}

@article{jia2018transfer,
  title={Transfer learning from speaker verification to multispeaker text-to-speech synthesis},
  author={Jia, Ye and Zhang, Yu and Weiss, Ron and Wang, Quan and others},
  journal={Proc. of NeurIPS},
  volume={31},
  year={2018}
}

@article{klein2024source,
  title={Source tracing of audio deepfake systems},
  author={Klein, Nicholas and Chen, Tianxiang and Tak, Hemlata and Casal, Ricardo and Khoury, Elie},
  journal={arXiv preprint arXiv:2407.08016},
  year={2024}
}

@article{kong2020diffwave,
  title={Diffwave: A versatile diffusion model for audio synthesis},
  author={Kong, Zhifeng and Ping, Wei and Huang, Jiaji and Zhao, Kexin and Catanzaro, Bryan},
  journal={arXiv preprint arXiv:2009.09761},
  year={2020}
}

@article{kong2020hifi,
  title={Hifi-gan: Generative adversarial networks for efficient and high fidelity speech synthesis},
  author={Kong, Jungil and Kim, Jaehyeon and Bae, Jaekyoung},
  journal={Proc. of NeurIPS},
  volume={33},
  pages={17022--17033},
  year={2020}
}

@inproceedings{liu2024groot,
  title={Groot: Generating robust watermark for diffusion-model-based audio synthesis},
  author={Liu, Weizhi and Li, Yue and Lin, Dongdong and Tian, Hui and Li, Haizhou},
  booktitle={Proc. of ACM MM},
  year={2024}
}

@article{wen2025sok,
  title={SoK: How Robust is Audio Watermarking in Generative AI models?},
  author={Wen, Yizhu and Innuganti, Ashwin and Ramos, Aaron Bien and Guo, Hanqing and Yan, Qiben},
  journal={arXiv preprint arXiv:2503.19176},
  year={2025}
}

@inproceedings{rix2001perceptual,
  title={Perceptual evaluation of speech quality (PESQ)-a new method for speech quality assessment of telephone networks and codecs},
  author={Rix, Antony W and Beerends, John G and Hollier, Michael P and Hekstra, Andries P},
  booktitle={Proc. of ICASSP},
  volume={2},
  pages={749--752},
  year={2001}
}

@article{taal2011stoi,
  title   = {An Algorithm for Intelligibility Prediction of Time-Frequency Weighted Noisy Speech},
  author  = {Taal, Cees H. and Hendriks, Richard C. and Heusdens, Richard and Jensen, Jesper},
  journal = {IEEE Transactions on Audio, Speech, and Language Processing},
  volume  = {19},
  number  = {7},
  pages   = {2125--2136},
  year    = {2011}
}

@inproceedings{reddy2021dnsmos,
  title     = {DNSMOS: A Non-Intrusive Perceptual Objective Speech Quality Metric to Evaluate Noise Suppressors},
  author    = {Reddy, Chandan K. A. and Gopal, Vishak and Cutler, Ross},
  booktitle = {Proc. of ICASSP},
  year      = {2021},
  pages     = {6493--6497}
}

@inproceedings{li2024proactive,
  author    = {Qi Li and Xiaodong Lin},
  title     = {Proactive Audio Authentication Using Speaker Identity Watermarking},
  booktitle = {PST},
  pages     = {1--10},
  year      = {2024}
}

\end{document}